\documentclass[noinfo,final,crop]{bioinfo}
\copyrightyear{2016}
\pubyear{2016}

\access{Advance Access Publication Date: Day Month Year}
\appnotes{Manuscript Category} 

\usepackage{ifpdf}
\usepackage{url}
\usepackage{tikz}
\usetikzlibrary{arrows}

\usepackage{tikz}
\usepackage{pgfplots}

\usepackage{pgffor}

\usepackage[ruled,vlined]{algorithm2e}

\makeatletter
\newcommand{\removelatexerror}{\let\@latex@error\@gobble}
\makeatother

\newcommand{\q}{\phantom0}
\newcommand{\qc}{\phantom{0,}}
\newcommand{\mcOT}{\multicolumn{5}{c}{\em out of time ($>12$ hours)}}

\DeclareMathOperator{\kmercount}{count}
\DeclareMathOperator{\weight}{weight}
\DeclareMathOperator{\prob}{prob}
\DeclareMathOperator{\rate}{rate}

\DeclareMathOperator{\tp}{TP}
\DeclareMathOperator{\tn}{TN}
\DeclareMathOperator{\fp}{FP}
\DeclareMathOperator{\fn}{FN}
\DeclareMathOperator{\sensitivity}{sensitivity}
\DeclareMathOperator{\precision}{precision}
\DeclareMathOperator{\gain}{gain}

\newcommand{\base}[1]{\mbox{\sffamily #1}}

\newif\iffigsinpdf

\begin{document}
\firstpage{1}

\subtitle{Sequence analysis} 

\title[RECKONER: Read Error Corrector Based on KMC]{RECKONER: Read Error Corrector Based on KMC}

\author[D{\l}ugosz and Deorowicz]{Maciej D{\l}ugosz\,$^{\text{\sfb 1}}$ and Sebastian Deorowicz\,$^{\text{\sfb 1,}*}$}
\address{$^{1}$Institute of Informatics, Silesian University of Technology, Akademicka 16, 44-100 Gliwice, Poland}

\corresp{$^\ast$To whom correspondence should be addressed.}

\history{Received on XXXXX; revised on XXXXX; accepted on XXXXX}

\editor{Associate Editor: XXXXXXX}

\abstract{\textbf{Motivation:} Next-generation sequencing tools have enabled producing of huge amount of
genomic information at low cost. Unfortunately, presence of sequencing
errors in such data affects quality of downstream analyzes.
Accuracy of them can be improved by
performing error correction. Because of huge amount of such data
correction algorithms have to: be fast, memory-frugal, and provide high
accuracy of error detection and elimination for
variously-sized organisms.\\
\textbf{Results:} We introduce a new algorithm for genomic data correction, capable of processing
eucaryotic 300\,Mbp-genome-size, high error-rated data using less than 4\,GB of RAM in less than
40\,minutes on 16-core CPU.
The algorithm allows to correct sequencing data at better or comparable level than competitors.
This was achieved by using very robust KMC~2 $k$-mer counter, new method of erroneous regions correction based on both $k$-mer counts and FASTQ quality indicators as well as careful optimization.\\
\textbf{Availability:} Program is freely available at \url{http://sun.aei.posl.pl/REFRESH}.\\
\textbf{Contact:} \href{sebastian.deorowicz@polsl.pl}{sebastian.deorowicz@polsl.pl}\\
}

\maketitle

\section{Introduction}

For several years we have been witnessing of amazing advances in developing of
DNA sequencing technologies. The famous Sanger method~\citep{SNS1977} 
has been superseded by so-called next-generation sequencing (NGS) technologies.
The instruments by Illumina/Solexa, Roche~454, Ion Torrent, ABI SOLiD~\citep{M10} allow producing huge amounts of sequencing reads at low cost.
This is, however, occupied by higher error rate.
The errors can be classified as: substitutions, which involves altering of
nucleotides by erroneous ones and indels, which involves insertions of nucleotide
sequences into another sequence or deletions of stretches of nucleotide sequences.
In Illumina, currently dominating technology, a major group of errors involve substitutions~\citep{LBM15}.

DNA sequencing data are used in various applications,
like \emph{de novo} assembly, reassembly, metagenomics, detecting of
single nucleotide polymorphisms (SNPs), gene expression analysis, to enumerate a few. 
The accuracy of the input data is crucial in all the cases.
Therefore, the correction of errors in reads is currently an important and popular issue.
The existing solutions are discussed and compared in the recent surveys: \citep{YCA13}, \citep{MI15}, \citep{LBM15}.
Yang \emph{et al.}\ classify the
correction algorithms into three groups: (\emph{i}) $k$-spectrum-based, (\emph{ii}) suffix-tree/array-based,
and (\emph{iii}) multiple-sequence-alignment-based.

%
%

The algorithms of the first category start from extracting all valid fragments of reads of length $k$ (called $k$-mers).
They suppose, that due to data redundancy, the majority of $k$-mers would have correspondent
(in sense, that it derives from the same fragment of the genome) $k$-mers also placed
in other reads. 
Then, rare $k$-mers are altered to the most similar frequent $k$-mers.
This, of course, means the correction of reads that contain the rare $k$-mers.
The algorithms tend to introduce the least possible number of changes, or, more precisely, such changes that repair most likely errors.
This category includes:
Quake \citep{KSS10}, RACER \citep{IM13}, BLESS \citep{HXCMH14}, Blue \citep{GDPB12}, Musket \citep{LSS13},
Lighter \citep{LFL14}, Trowel \citep{LMHHKW14}, Pollux \citep{MBM15}, BFC \citep{Li2015},
Ace \citep{SR15}.

The suffix-tree/array-based algorithms, like SHREC \citep{SSP09}, HiTEC \citep{IFI11}
also extract $k$-mers, but they utilize simultaneously
different values of $k$ and store $k$-mer sets in a suffix data structure.

The multiple-sequence-alignment-based solutions choose from the input dataset such reads
that seem to origin from the same fragment of genome. Then they perform one of
multiple sequence alignment algorithms on these reads with aim of finding
the consensus form of the reads. 
This category involves Coral \citep{SS11}, ECHO \citep{KCS11}, Karect \citep{AKS15}.
Fiona \citep{SWHDNRR14} is a hybrid approach, as it uses suffix tree and also performs multiple sequence alignment.

The read correction problem is hard due to the following reasons.
The ``source'' genome is not known in advance, so determining the proper
form of a read could be achieved only with heuristic methods. 
Of course, especially in a case of low coverage, the algorithm can sometimes change the correct symbol to the wrong one.
The repeats typical especially for large genomes, could also cause problems with
choosing proper correction from a set of a few possibilities. 

The most important feature of the read corrector is of course the quality of the results, but we should remember that the amount of input data is huge, as the number of reads can be counted in hundreds of millions.
This can lead to large memory occupation if the algorithm constructs a complex data structure from all the reads.
The computation time is also an important factor.
The above-described reasons motivate deployment of new solutions in this field.

We present a new algorithm for correction of read errors---RECKONER. Our solution is able to correct
eucaryotic 300~Mbp sequencing data using less than 4~GB of RAM in less than 40~minutes
on a machine equipped with 16~CPU~cores,
providing correction accuracy better or comparable to competitive methods.
The presented results of evaluation of our and state-of-the-art algorithms are 
on both real and simulated data, including assay of influence of correction on various
statistical indicators of correction quality and on results of typical applications of
NGS reads.

RECKONER performs $k$-mer counting with an efficient KMC~2 algorithm~\citep{DKGD15} and stores
the $k$-mer database, required during the correction phase, in a compact data structure provided by
KMC API. It also improves previous methods of error detection by more intensive utilization of base
quality indicators. RECKONER introduces a new method of rating possible corrections based on
both read bases quality indicators and $k$-mer counts. The algorithm performs correction in parallel and
allows to process the compressed input data.

\begin{methods}
\section{Methods}
\subsection{General idea}

The general idea of $k$-spectrum-based algorithms is similar. First, they perform
$k$-mer counting, i.e., counting a number of appearances of every substring of length~$k$
present in the input data. 
The obtained counters are used to determine whether the specified fragment of a read is correct or not. It relies on the
observation, that in NGS technologies data redundancy causes, that with reasonably high probability
every short fragment of a genome (i.e., with length $k$) would appear multiple times in the input data.

As RECKONER is a $k$-spectrum based algorithm, its workflow is in the high level similar to other algorithms from this family.
It consists of four
stages:
\begin{enumerate}
\item $k$-mer counting, 
\item determining threshold of number of $k$-mer appearances, which indicates \emph{trusted} and \emph{untrusted} $k$-mers, 
\item removing \emph{untrusted} $k$-mers from the $k$-mer database,
\item correcting the reads. 
\end{enumerate}
This solution relies on altering of possibly
erroneous bases in order to achieve a read variant which is, with the highest probability, the correct one.
The main part of RECKONER
is based on BLESS version 0.12, which provided a spine of error correction scheme,
although we have complemented it by series of improvements.

The main initial changes are employing KMC~2~\citep{DKGD15} for $k$-mer counting and storing
$k$-mers together with counters in the KMC database accessible with KMC~API.
Such solution has allowed us to prepare a new method of rating corrections, which
utilizes both read quality indicators and $k$-mer counters.
Moreover, we have improved the idea of read extension while performing corrections near the read ends. 
RECKONER uses not only the information of extension success but also about the extension quality.
We also improved utilization of base quality indicators. The idea is to
use poor quality values to indicate places, which should be
checked more accurately while determining proper read form.

RECKONER is strongly time- and memory-optimized and is parallelized with OpenMP.

Meanwhile, BLESS authors released new versions
of their software, which were independently supplemented by various novel functions, including
$k$-mer counting with KMC~2 and algorithm parallelization. 
Nevertheless, the other improvements introduced in RECKONER are not present in the new release of BLESS.

\subsection{$k$-mer counting and cutoff}
The first stage of RECKONER is counting the appearances of every $k$-mer in all input FASTQ files.
It is performed by KMC~2 in non-quality-aware mode. Counting does not distinguish between $k$-mers and their
reversed compliments, i.e., KMC counts only canonical $k$-mers. Canonical $k$-mer is defined as lexicographically
smaller of a $k$-mer and its reverse compliment. 
This approach reduces memory requirements nearly twice (compared to counting all the $k$-mers as they appear in the reads), which is possible since there is no information in the input data
about direction of the strand the corrected fragment originates from. Then, an auxiliary tool, called \emph{cutter},
creates $k$-mer appearances histogram, i.e., histogram of numbers of $k$-mers, which occur in
the input data particular number of times. 
The histogram is used to find the threshold, being a number of $k$-mer appearances used to qualify a $k$-mer into either group of \emph{trusted}
(error-free, in BLESS -- \emph{solid}) or \emph{untrusted} (erroneous, \emph{weak}) ones.

The method of determining the threshold is based on the one implemented in BLESS. It exploits the observation,
that untrusted $k$-mers appear in the input data generally rarely (like one or two times). On the other hand, in NGS technologies trusted
$k$-mers, providing the adequate input data coverage, have frequencies
significantly larger (e.g., ten or more), so one can expect, that
there is a region between low-coverage and high-coverage $k$-mers with $k$-mers with middle-sized frequencies.
RECKONER looks for the first minimum in the histogram (i.e., the first value $x$ such that both the number of $k$-mers appearing $x-1$ times and $x+1$ times is larger than the number of $k$-mers appearing $x$ times).
This value is chosen as the threshold but to prevent from 
choosing too large number, which can be caused by not unified genome coverage,
the threshold is upper-bounded by~$5$.\looseness=-1

Then, KMC database is truncated by removing $k$-mers appearing fewer times than the threshold.
This reduces the memory consumption during the next stages, as the database of $k$-mers is stored in the main memory.

Read correction algorithms differ in how they count the $k$-mer counters.
Sometimes they use 	external tools like Jellyfish~\citep{MK11} in Quake, KMC~2 in RECKONER.
The chosen method has great impact on the processing speed and memory consumption of this stage.
The algorithms also differ in the data structure used for storing $k$-mers (sometimes together with the related counters).
BLESS stores such data in a Bloom filter, Quake uses bitmap of size of the entire space of $k$-mers.
The side effect of these decisions is that both BLESS and Quake cannot utilize the $k$-mer counters in the correction phase, as they only know whether the $k$-mer is trusted or not.
Sometimes even this knowledge is imperfect (e.g., Bloom filters allow for some number of false positive answers of appearance of $k$-mers, although this problem has been partially solved in BLESS).\looseness=-1

RECKONER uses $k$-mer database produced by KMC~2, accessible by KMC~API.
This way it can relay not only on the presence of $k$-mers but also on the exact number of appearances of each $k$-mer in the input dataset.
KMC~2 allows to calculate the quality-aware counters taking into account
base qualities (``$q$-mers'') \citep{KSS10}.
Therefore, we experimented with using them in the early versions of RECKONER, but as they have not bring a significant improvement in the quality of a correction, we finally discarded this approach.

\begin{figure}
\centering
\begin{tikzpicture}[>=stealth, x=0.45cm, y=0.4cm]
\footnotesize
\draw(8.5,20.5)[black] node{\sffamily Input file};
\draw[->, thick] (8.5, 20) -- (8.5, 19.2);

\draw(2.8,15.5)[black] node{\sffamily KMC database};
\draw(14.2,15.5)[black] node{\sffamily KMC database};
\draw[->, thick] (2.8, 15) -- (2.8, 14.2);
\draw[->, thick] (14.2, 15) -- (14.2, 14.2);

\filldraw[black!25!white](6,18) rectangle(11,19);
\draw[thick](6,18) rectangle(11,19);
\draw(8.5,18.5)[black] node{\sffamily Chunkifying};
\draw[->, thick] (8.5, 18) -- (10.75, 17.2);

\filldraw[black!10!white](6,16.5) rectangle(11,17);
\draw[thick](6,16.5) rectangle(11,17);
\draw[thick](6.5,16.5)--(6.5,17);
\draw[thick](7.0,16.5)--(7.0,17);
\draw[dotted, thick](7.25,16.75)--(9.75,16.75);
\draw[thick](10.0,16.5)--(10.0,17);
\draw[thick](10.5,16.5)--(10.5,17);
\draw[right](11.25,16.75) node{\sffamily Set of chunks};
\draw[->, thick] (6.25, 16.5) -- (3.5, 14.2);
\draw[->, thick] (6.25, 16.5) -- (13.5, 14.2);

\filldraw[black!25!white](1,13) rectangle(6,14);
\draw[thick](1,13) rectangle(6,14);
\draw(3.5,13.5)[black] node{\sffamily Thread $1$};
\draw[->, thick] (3.5, 13) -- (3.5, 12.2);

\filldraw[black!25!white](1,11) rectangle(6,12);
\draw[thick](1,11) rectangle(6,12);
\draw(3.5,11.5)[black] node{\sffamily Error detection};
\draw[->, thick] (3.5, 11) -- (3.5, 10.2);

\filldraw[black!25!white](1,9) rectangle(6,10);
\draw[thick](1,9) rectangle(6,10);
\draw(3.5,9.5)[black] node{\sffamily Greedy correction};
\draw[->, thick] (3.5, 9) -- (3.5, 7.2);

\filldraw[black!25!white](1,6) rectangle(6,7);
\draw[thick](1,6) rectangle(6,7);
\draw(3.5,6.5)[black] node{\sffamily Path rating};
\draw[->, thick] (3.5, 6) -- (6.15, 3.2);

\filldraw[black!25!white](11,13) rectangle(16,14);
\draw[thick](11,13) rectangle(16,14);
\draw(13.5,13.5)[black] node{\sffamily Thread $t$};
\draw[->, thick] (13.5, 13) -- (13.5, 12.2);

\filldraw[black!25!white](11,11) rectangle(16,12);
\draw[thick](11,11) rectangle(16,12);
\draw(13.5,11.5)[black] node{\sffamily Error detection};
\draw[->, thick] (13.5, 11) -- (13.5, 10.2);

\filldraw[black!25!white](11,9) rectangle(16,10);
\draw[thick](11,9) rectangle(16,10);
\draw(13.5,9.5)[black] node{\sffamily Greedy correction};
\draw[->, thick] (13.5, 9) -- (13.5, 7.2);

\filldraw[black!25!white](11,6) rectangle(16,7);
\draw[thick](11,6) rectangle(16,7);
\draw(13.5,6.5)[black] node{\sffamily Path rating};
\draw[->, thick] (13.5, 6) -- (6.35, 3.2);

\draw[dotted,thick](6.5, 13.5) -- (10.5, 13.5);

\draw[dotted,thick](6.5, 11.5) -- (10.5, 11.5);

\draw[dotted,thick](6.5, 9.5) -- (10.5, 9.5);

\draw(8.5,8)[black] node{\sffamily Correction paths};

\draw[dotted,thick](6.5, 6.5) -- (10.5, 6.5);

\draw(8.5,5.25)[black] node{\sffamily Best corrections};

\filldraw[black!10!white](6,2.5) rectangle(11,3);
\draw[thick](6,2.5) rectangle(11,3);
\draw[thick](6.5,2.5)--(6.5,3);
\draw[thick](7.0,2.5)--(7.0,3);
\draw[dotted, thick](7.25,2.75)--(9.75,2.75);
\draw[thick](10.0,2.5)--(10.0,3);
\draw[thick](10.5,2.5)--(10.5,3);
\draw[right](11.25,2.5) node{\sffamily Corrected chunks};
\draw[->, thick] (10.75, 2.5) -- (8.5, 1.2);

\filldraw[black!25!white](6,0) rectangle(11,1);
\draw[thick](6,0) rectangle(11,1);
\draw(8.5,0.5)[black] node{\sffamily Result integrating};
\draw[->, thick] (8.5, 0) -- (8.5, -0.8);

\draw(8.5,-1.5)[black] node{\sffamily Corrected file};

\end{tikzpicture}

\caption{Correction process (first $k$-mer correction is disregarded)}
\label{fig:correctflow}
\end{figure}
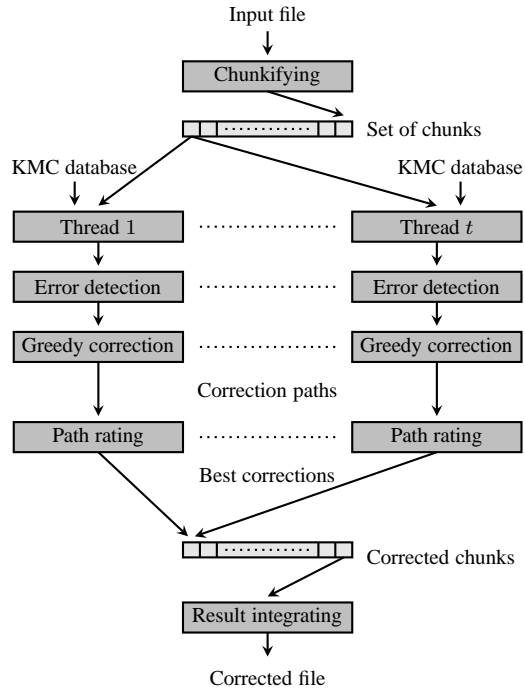
\begin{figure}

\begin{tikzpicture}[>=stealth, x=1.25cm, y=0.45cm]

\draw(-1.5, 2) node [scale=0.5]{\large Bases' indices};
\draw(-1.5, 1) node [scale=0.5]{\large Bases' qualities};
\draw(-1.5, 0) node [scale=0.5]{\large Bases};
\draw(-1.5, -3.25) node [scale=0.5]{\large $k$-mers counts};
\draw(-1.5, -6.2) node [scale=0.5]{\large Path $1$};
\draw(-1.5, -10.3) node [scale=0.5]{\large Path $2$};

\foreach \n in {0,...,7}
{
\draw(0.2 * \n, 2) edge [dotted](0.2 * \n, -1.0 - 0.3 * \n)[black!10];
}

\foreach \n in {0,...,6}
{
\draw(1.6 + 0.2 * \n, 2) edge [dotted](1.6 + 0.2 * \n, -11.3 - 0.3 * \n)[black!10];
}

\foreach \n in {0,...,6}
{
\draw(\n * 0.2 + 3, 2) edge [dotted](0.2 * \n + 3, -13.4)[black!10];
}

\foreach \n in {79,...,99}
{
\draw({(\n - 79) * 0.2}, 2) node [scale=0.5, inner sep=0, anchor=west] {\ttfamily\fontsize{9pt}{12pt}\selectfont \n};
}

\newcounter{i}

\setcounter{i}{0}
\foreach \n in {39,40,40,40,40,40,40,40,40,40,40,40,39,27,18,9,9,8,8,17,43}
{
\draw(\value{i} * 0.2, 1) node [scale=0.5, inner sep=0, anchor=west] {\ttfamily\fontsize{9pt}{12pt}\selectfont \n};
\addtocounter{i}{1}
}

\draw(0, 0) node [inner sep=0, anchor=east] {\ttfamily \ldots};
\setcounter{i}{0}
\foreach \n in {A,T,T,A,T,A,A,A,C,G,C,A,T,C,T,A,T,C,G,A,C}
{
\draw(\value{i} * 0.2, 0) node [inner sep=0, anchor=west] {\ttfamily \n};
\addtocounter{i}{1}
}

\setcounter{i}{0}
\foreach \n in {30,26,25,16,18,24,32,35,0,0,0,0,0,0,29}
{
\draw(-0.05 + 0.2 * \value{i}, -1.0 - 0.3 * \value{i}) node [scale=0.5, anchor=east, inner sep=0.05]{\fontsize{9pt}{12pt}\selectfont \n};
\draw (0.2 * \value{i}, -1.0 - 0.3 * \value{i}) edge  (1.4 + 0.2 * \value{i}, -1.0 - 0.3 * \value{i});
\addtocounter{i}{1}
}


\draw (-1.5,-5.7) edge [black!20](4.5,-5.7);

\setcounter{i}{0}
\foreach \n in { , , , , , , , , , , , , , ,A, , , , ,T, }
{
\draw(\value{i} * 0.2, -6.2) node [inner sep=0, anchor=west] {\ttfamily \n};
\addtocounter{i}{1}
}

\setcounter{i}{0}
\foreach \n in {30,24,27,17,35,21,20,22}
{
\draw({-0.05 + 0.2 * (\value{i} + 8)}, -7.2 - 0.3 * \value{i}) node [scale=0.5, anchor=east, inner sep=0.05]{\fontsize{9pt}{12pt}\selectfont \n};
\draw ({0.2 * (\value{i} + 8)}, -7.2 - 0.3 * \value{i}) edge ({1.4 + 0.2 * (\value{i} + 8)}, -7.2 - 0.3 * \value{i});
\addtocounter{i}{1}
}
\draw(0.2 * 15 - 0.5, -7.2 - 0.3 * 7) node [scale=0.5, anchor=east]{\large Best extending $k$-mer};


\draw (-1.5,-9.8) edge [black!20](4.5,-9.8);

\setcounter{i}{0}
\foreach \n in { , , , , , , , , , , , , , ,A,A, ,C, , , }
{
\draw(\value{i} * 0.2, -10.3) node [inner sep=0, anchor=west] {\ttfamily \n};
\addtocounter{i}{1}
}

\setcounter{i}{0}
\foreach \n in {30,25,37,29,45,28,23,24}
{
\draw({-0.05 + 0.2 * (\value{i} + 8)}, -11.3 - 0.3 * \value{i}) node [scale=0.5, anchor=east, inner sep=0.05]{\fontsize{9pt}{12pt}\selectfont\n};
\draw ({0.2 * (\value{i} + 8)}, -11.3 - 0.3 * \value{i}) edge ({1.4 + 0.2 * (\value{i} + 8)}, -11.3 - 0.3 * \value{i});
\addtocounter{i}{1}
}
\draw(0.2 * 15 - 0.5, -11.3 - 0.3 * 7) node [scale=0.5, anchor=east]{\large Best extending $k$-mer};

\end{tikzpicture}

\caption{Example of correction paths rating}
\label{fig:example}
\end{figure}
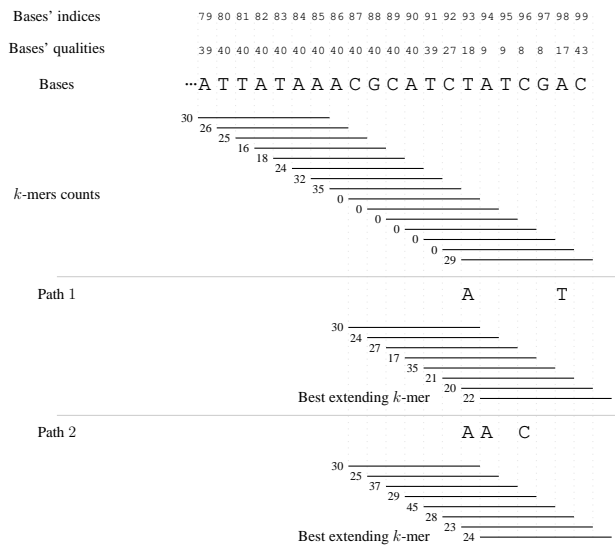

\subsection{Correcting of the reads}
\subsubsection*{Choice of erroneous read regions}
The last stage of the processing, the read correction, consists of a few steps
applied separately for every FASTQ file (Fig. \ref{fig:correctflow}):
(\emph{i}) read scanning and partitioning (\emph{chunkifying}) for multithreading,
(\emph{ii}) detecting of the erroneous regions, (\emph{iii}) generating and rating
of error correction paths, and (\emph{iv}) results integrating.

The idea of read scanning and its parameters are similar to corresponding step of BLESS.
%
It relies on reading every input file with aim of dividing it
into parts (\emph{chunks}) processed separately by different threads.

Every correction thread processes a subset of chunks appointed in the previous step. First, every read
is split into regions that are marked as either proper or  erroneous. Erroneous
regions are determined by extracting its $k$-mers and by checking their presence in the KMC database.
If a current $k$-mer is absent from the database, it makes a suggestion, that it
originates from a read region having at least one erroneous base.
Erroneous regions are expanded if the adjacent bases have low qualities (i.e., smaller than $10$).
Moreover, both type regions can be removed if their lengths are too small (less than $2$).

Fig.~\ref{fig:example} shows an example of the process of finding erroneous regions.
RECKONER extracts a set of $k$-mers from an exemplary read and obtains their counters from the KMC database.
The last $k$-mer has counter $29$ despite of presence of low-quality bases within. 
Next six
$k$-mers (from right to left) have number of appearances $0$ (they have been removed from the database).
The presence of such $k$-mers causes that RECKONER marks as an erroneous region of indices from $93$ to $98$.
Then, it expands this region to base $99$, to remove ``proper'' region with length
shorter than $2$ bases.\looseness=-1

\subsubsection*{Error correction and correction rating}
Next step consists of introducing changes into reads (Fig. \ref{fig:example}).
The method of correction of detected wrong bases exploits greedy algorithm similar to competitive solutions, like BLESS and Quake.
The essential part of the algorithm is described below. Every base in an erroneous region,
respectively from right to left (i.e., 3' to 5' end -- situation \emph{1}), for the erroneous region situated in the 5' end, or
from left to right (i.e., 5' to 3' end -- situation \emph{2}), for the other regions, can be altered to another
one until the adequate $k$-mer with the base become trusted; low-quality bases are changed even if
the $k$-mer is already trusted. Especially, situation \emph{2} includes erroneous regions situated between two proper regions.

Following its definition, it can be expected, that the erroneous region is of length at least~$k$. 
If it is shorter
or there is no proper region in the read,
 then correction of the first $k$-mer
is performed and after successful result of this correction the standard correction in situation \emph{2}
is performed. The correction of the first $k$-mer relies on altering bases with quality indicators
smaller than $10$ or, in case of failure, by altering every single base in the $k$-mer.

Every base present in the erroneous region is a candidate to be modified, however,
modifications are introduced only to bases satisfying at least one of two conditions:
(\emph{i}) the base makes the trailing (situated contrariwise to the correction direction) $k$-mer untrusted or
(\emph{ii}) it has a low value of its quality indicator (accompanying the base in the read).

The second condition is introduced in RECKONER.
The idea is to prevent a situation when some read region contains a large number of low-quality bases,
strongly suggesting that some of them are erroneous, even if the $k$-mers extracted from the region are trusted.

Finally, many candidate solutions can be obtained. The next step
is paths rating to choose the best one.


The exemplary read would be corrected by RECKONER starting from base $93$. RECKONER would try to change the base unless
$k$-mer starting at $87$th position will become trusted. Let us suppose, that the only
possibility is to change it to~\base{A}. Then, it checks whether $k$-mer starting at $88$th position become trusted. If yes, it
continues correcting following bases until region's $k$-mers become trusted. Let us suppose, that the only
possibility is to change the $98$th base to~\base{T}. 
These two changes constitute the first path of changes.

When RECKONER detects a base with low quality (smaller than $10$), which has not
to be changed to cause a current $k$-mer become
trusted, it treats ``no change'' possibility as the beginning of one correction path, and regardless of trust of the $k$-mer tries to change the base;
if the change causes, that $k$-mer is still trusted, then the change starts the next correction path.
Let us suppose, that base $94$ can be changed to~\base{A} to cause (regardless of being trusted after changing base $93$),
that $k$-mer starting from the $88$th position remains
trusted. This way we have started the second correction path. Let us suppose, that to cause the entire
erroneous region to be trusted, we also have to change base $96$ to~\base{C}.

Like BLESS, RECKONER performs read extending. 
If a corrected base is located nearly ends of the read, there is no possibility to extract sufficient number
of $k$-mers the base builds. The solution relies on checking also symbols, that possibly would be
a continuity of the read in aim of finding such combination of that symbols, that could constitute
trusted $k$-mers with symbols on the specified end of the read.
If for a specified correction path there is no possibility to
find extending symbols then the current correction the path is rejected.

Correction of bases with greedy algorithm would cause in rare situations
extreme increase of a number of considered paths of correction. Because of that, we have introduced limitations on the
maximum number of changes performed in a considered region of a read. If the limit is reached,
searching of paths for the region is interrupted.

In case of finding multiple paths of correction
there is a need to rank them. This rating is based on the following formula.
Let us define read~$r$ as a sequence over the alphabet $\Sigma=\{\base{A}, \base{C}, \base{G}, \base{T}\}$ of length $\ell$.
If the sequencing data contains another symbols, then they are altered to \base{A}.
Let $r[s]$ denotes the $s$-th symbol of $r$, $-1 \leq s \leq \ell$, where
$r[0] r[1] \ldots r[\ell-1]$ are symbols originating from the input read;
$r[-1]$ and $r[\ell]$ are symbols belonging to the read extension (if it includes that symbols).
Moreover, $r[a, b]$ means $r[a] r[a+1]\ldots r[b]$.

Let us define $p$ as a sequence of length $\ell$ of probabilities, that the particular symbols of
the corresponding sequence $r$ are incorrect. The notation $p[t]$ means the $t$-th value of $p$, $0 \leq t \leq \ell-1$.

Let us suppose, that $r[a, b]$ is an erroneous region of~$r$, where $0 \leq a < b \leq \ell-1$.
Moreover, in situation \emph{1} we know that $b < \ell-k+1$ and in situation \emph{2} that $a \geq k-1$.
Let us denote as $m$ index of the left-most (\emph{1}) or the right-most (\emph{2}) modified base.
The distance $d$ of $m$ from the read's end is:
\begin{equation*}
d=\begin{cases}
m & \text{(\emph{1}),}\\
\ell - m - 1 & \text{(\emph{2}).}
\end{cases}
\end{equation*}
The number $e$ of bases extending the read is determined as:
\begin{equation*}
e=\begin{cases}
0 & \text{ if } d \geq k \text{,}\\
\min(k-d-1, 5) & \text{otherwise.}
\end{cases}
\end{equation*}
By set $K_\text{cov}$ of $k$-mers covering a region $r[a, b]$ we mean:
\begin{equation*}
K_\text{cov}=\begin{cases}
\{r[a, a+k-1],\ldots, r[b, b-k+1]\} & \text{(\emph{1}),}\\
\{r[a-k+1, a],\ldots, r[b+k-1, b]\} & \text{(\emph{2}).}
\end{cases}
\end{equation*}
The first extending $k$-mer $x_\text{ext}$ is defined as:
\begin{equation*}
x_\text{ext}=\begin{cases}
r^*[-1, k-2] & \text{(\emph{1}),}\\
r^*[\ell-k+1, \ell] & \text{(\emph{2}).}
\end{cases}
\end{equation*}
and, finally, the set $K^*$ of rating $k$-mers is defined as:
\begin{equation*}
K^*=\begin{cases}
K^*_\text{cov} & \text{ if } e=0 \text{,}\\
K^*_\text{cov} \cup x_\text{ext} & \text{ if } e>0 \text{.}
\end{cases}
\end{equation*}
where $K^*_\text{cov}$ is an equivalent of $K_\text{cov}$ containing $k$-mers
taken from a modified read $r^*$, i.e., the version of~$r$ with applied modifications according to the currently evaluated correction path.

If there are more than one possible (trusted) first extending $k$-mer,
$K^*$~contains the one with the biggest number of appearances 
(in a tie we choose any of them, since here only the $k$-mer counter is important).

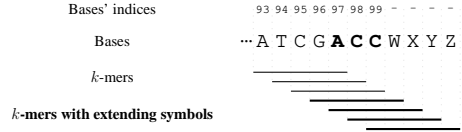
\begin{figure}

\begin{tikzpicture}[>=stealth, x=1.25cm, y=0.42cm]

\draw(-1.5, 2) node [scale=0.5]{\large Bases' indices};
\draw(-1.5, 1) node [scale=0.5]{\large Bases};
\draw(-1.5, -0.15) node [scale=0.5]{\large $k$-mers};
\draw(-1.5, -1.35) node [scale=0.5]{\large \textbf{$k$-mers with extending symbols}};

\foreach \n in {0,...,6}
{
\draw(0.2 * \n, 2) edge [dotted](0.2 * \n, 0 - 0.3 * \n)[black!10];
}

\foreach \n in {0,...,4}
{
\draw(\n * 0.2 + 1.4, 2) edge [dotted](0.2 * \n + 1.4, -1.8)[black!10];
}

\foreach \n in {93,...,99,-,-,-,-}
{
\draw({(\n - 93) * 0.2 + 0.1}, 2) node [scale=0.5, inner sep=0] {\ttfamily\fontsize{9pt}{12pt}\selectfont  \n};
}

\foreach \n in {100,...,103}
{
\draw ({(\n - 93) * 0.2 + 0.1}, 2) node [scale=0.5, inner sep=0] {\ttfamily -};
}

\newcounter{j}

\draw(0, 1) node [inner sep=0, anchor=east] {\ttfamily \ldots};
\setcounter{j}{0}
\foreach \n in {A,T,C,G,\textbf{A},\textbf{C},\textbf{C},W,X,Y,Z}
{
\draw(\value{j} * 0.2 + 0.1, 1) node [inner sep=0] {\ttfamily \n};
\addtocounter{j}{1}
}

\setcounter{j}{0}
\foreach \n in {0,...,2}
{
\draw (0.2 * \value{j}, 0.0 - 0.3 * \value{j}) edge (1.0 + 0.2 * \value{j}, 0.0 - 0.3 * \value{j});
\addtocounter{j}{1}
}

\foreach \n in {3,...,6}
{
\draw [thick] (0.2 * \value{j}, 0.0 - 0.3 * \value{j}) edge (1.0 + 0.2 * \value{j}, 0.0 - 0.3 * \value{j});
\addtocounter{j}{1}
}

\end{tikzpicture}

\caption{$5$-mers covering bases 97 -- 99; \base{W}, \base{X}, \base{Y}, \base{Z} -- any extending symbols (at most $5$)}
\label{fig:covering}
\end{figure}

For rating we use only the first extending $k$-mer, however, the necessary condition for acceptance
and rating of a path is that every $x \in K^*$, is trusted and
at least one $k$-mer built with every of extending (at most $5$) positions is trusted (Fig. \ref{fig:covering}).

Let $\kmercount(x)$ be the number of $k$-mer $x$ appearances and $\weight$ be defined as follows:
\begin{equation*}
\weight(x)=\begin{cases}
1 &\text{if }x \text{ contains no extending bases,}\\
0.5 &\text{if one base of }x \text{ is the extending base.}
\end{cases}
\end{equation*}
We also need $\prob$ function defined as follows:
\begin{equation*}
\prob(i)=\begin{cases}
p[i] &\text{if the }i\text{-th symbol has been changed,} \\
1 &\text{otherwise.} \end{cases}
\end{equation*}
Then, the region's correction rate is:
\begin{equation*}
\rate(a, b, K^*)={\left(\sum\limits_{x \in K^*}^{}\weight(x)\kmercount(x)\right)\left(\prod\limits_{i=a}^{b}\prob(i)\right) \over \sum\limits_{x \in K^*}^{}\weight(x)}.
\end{equation*}
Finally, the correction with the biggest rate is applied to the erroneous region.



The base qualities and finally base error probability, can be obtained from the input FASTQ file
and the number of $k$-mer appearances are taken from the KMC database. The idea of using base probabilities
for rating of correction paths has been previously proposed in Quake \citep{KSS10}, but there the probabilities have
been processed as a parameters of a nonparametric regression.
The advantage of the proposed method is utilizing both $k$-mer counts and read base qualities for corrections rating.
Particular erroneous regions are rated separately.

\begin{table}[!h]
\processtable{Bases qualities with corresponding probabilities.
\label{tab:qual}}
{
\renewcommand{\tabcolsep}{1.0em}
\begin{tabular}{@{\extracolsep{1.45em}}rr|rr}
\toprule
Quality & Probability & Quality & Probability\\
\midrule
18 & 0.015849 & 8  & 0.158489 \\
9  & 0.125893 & 17 & 0.019953 \\
\botrule
\end{tabular}
}{}
\end{table}

Based on Table \ref{tab:qual}, RECKONER would determine the rates for the exemplary paths $1$ and~$2$ as: 
%
%
{
\begin{eqnarray*}
\rate_1 &=& \frac{(30+24+27+17+35+21+20+0.5 \cdot 22)}{1+1+1+1+1+1+1+0.5} \cdot\\
&\cdot& (0.015849 \cdot 0.019953) =  0.0078\\
\rate_2 &=&\frac{(30+25+37+29+45+28+23+0.5 \cdot 24)}{ 1+1+1+1+1+1+1+0.5} \cdot\\
&\cdot& (0.015849 \cdot 0.125893 \cdot 0.158489) = 0.009656
\end{eqnarray*}}
so in the shown situation
RECKONER would apply path $2$ to the input read.

\section{Results}
We have performed a series of tests on both simulated and real data for
various-sized genomes, different sequencing coverages, read lengths, and data qualities.
These allowed us to compare the examined algorithms
from different points of view: correction quality characterized respectively by
statistical measures (simulated data), impact on \emph{de novo}
assembly and reassembly (real data), memory requirements, and time consumption. For testing we picked
data from different Illumina sequencers.

We compared RECKONER and the following state-of-the-art algorithms: Ace~\citep{SR15}, BFC~\citep{Li2015}, BLESS~\citep{HXCMH14},
Blue~\citep{GDPB12}, Karect~\citep{AKS15}, Lighter~\citep{LFL14}, Musket~\citep{LSS13}, Pollux~\citep{MBM15}, RACER~\citep{IM13}, Trowel~\citep{LMHHKW14}. 
We excluded from this group
the popular Quake due to its poor results according to the recent studies~\citep{HXCMH14,LMHHKW14,LSS13,MBM15}.

The experiments were performed on a computer equipped with 128~GB of RAM, four AMD Opteron 8378 processors, running under Fedora 16 x86-64 OS.
RECKONER was written in C++ and parallelized using OpenMP 3.1 library.
For compilation we used G++ 4.7.

\subsection{Real data evaluation}
In practice, the sequenced reads are mapped (reassembled) or \emph{de novo} assembled.
We performed an evaluation of correction algorithms on real data by assaying an impact of the correction
on results in both applications as they are sensitive to sequencing errors presence.
The tests were performed for three datasets described in Table~\ref{tab:datasets:real}.
Two of this datasets were taken from the survey by Molnar and Ilie~(\citeyear{MI15}).


\begin{table}[!t]
\processtable{Real datasets used in the experiments
\label{tab:datasets:real}}
{
\renewcommand{\tabcolsep}{0.5em}
\begin{tabular}{@{\extracolsep{0.45em}}lcccc}
\toprule
Organism & Accession no. & Genome len.		& No.\ of reads	& Read length\\
\midrule
\emph{S.\ cerevisiae} & ERR422544  & \q12.3\,Mbp &	\q4.78\,M	&	100\,bp	\\
\emph{C.\ elegans} & SRR543736  & 102.3\,Mbp 	&	57.72\,M	& 	101\,bp\\
\emph{M.\ acuminata} & ERR204808 & 472.2\,Mbp 	&	67,18\,M	&	108\,bp	\\

\botrule
\end{tabular}
}{}
\end{table}

\subsubsection*{\emph{De novo} assembly}
\emph{De novo} assembly is a process of building contiguous sequences of nucleotides, called \emph{contigs},
typically by detection of overlapping fragments of reads \citep{KSS10}.

In genome assembly finding of false overlaps or missing true overlaps can occur due to sequencing errors.
It may cause generation of false contigs, emergence of ambiguity or breaking of true contigs.
Moreover, many of assemblers model contig fragments and overlaps between them with a de Bruijn graph.
De Bruijn graph memory requirements strongly depend on a number of distinct $k$-mers in the input data. Every additional (e.g., erroneous) $k$-mer
causes at most $k$ faulty nodes appearing in the graph, what significantly impacts on memory requirement of assembly.

The quality of assembly was evaluated using several measures, i.e., NG50, NA50, N50.
In the main text the $\text{NG50}$ measure (such contig length, that contigs of length NG50 or more consist 50\,\% of the full genome) is used.
The other results (together with time of assembly and assembly memory requirements are given in Supplementary material).
For experiments we used Velvet assembler~\citep{ZB08}. $\text{NG50}$ values were determined with Quast~\citep{GGSV13}.

The results are given in Table~\ref{tab:results:de_novo}.
As it can be seen the best results are obtained by Karect, but the second place is for RECKONER.
What is important, only Karect, RECKONER, and Blue are always ranked among first 5 positions. 
The other algorithms perform poorly for at least one dataset.
Two of them perform even poorer than when no correction is used.\looseness=-1

\begin{table}[!t]
\processtable{Results of \emph{de novo} assembly
\label{tab:results:de_novo}}
{
\renewcommand{\tabcolsep}{0.5em}
\begin{tabular}{@{\extracolsep{0.45em}}lccc}
\toprule
&	\multicolumn{3}{c}{Organism}\\\cline{2-4}
Corrector & \emph{S.\ cerevisiae} & \emph{C.\ elegans} & \emph{M.\ acuminata} \\
\midrule
Karect				&\bf18,034	&	2,874	&\bf1,469\\	
RECKONER				&	17,800	&\bf2,973& 	1,384\\	
Blue					&	16,991	&	2,952	&	1,423\\	
Pollux				&	16,304	&	2,900	&	1,423\\	
BFC					&	17,292	&	2,731	&	1,342\\	
BLESS					&	16,861	&	2,783	&	1,332\\	
Lighter				&	16,167	&	2,848	&	1,318\\	
Musket				& 	16,639	&	2,562	&	1,279\\	
RACER					&	17,292	&	2,040	&	1,092\\	
\it Without corr.	&\it16,204	&\it2,521&\it1,335\\	
Ace					&	16,574	&	1,294	&	\qc913\\	
Trowel				& 	13,831	&	1,626	&	1,261\\	
\botrule
\end{tabular}
}{The given numbers are NG50 values obtained by Velvet assembler for data corrected using the examined correctors.
The correctors are ordered according to average rank.
The best values are in bold
}
\end{table}

\subsubsection*{Reassembly}
The goal of reassembly is to detect reads' locations in the genome by aligning them
to the reference genome. It cannot be done by simple searching of each read in
the reference genome. Intraspecific diversity, especially dissimilarities of single nucleotides
between genomes of the same species (single nucleotide polimorphisms), enforce use of more
sophisticated algorithms. In this case sequencing errors cause ambiguities of reads
matching and finding false matchings or missing true matchings.

We performed the evaluation using Bowtie~2~\citep{LS12}. As a measurement indicator we picked both number of modifications required to be introduced to align a read and number of reads successfully aligned to the genome.
Table~\ref{tab:results:mapping} presents the fractions of reads mapped by Bowtie~2 after correction.
Other results, i.e., time of mapping, memory consumption of Bowtie~2, fractions of reads mapped uniquely, fractions of reads mapped with 1, 2, \ldots, 5 and more mismatches are given in Supplementary material.
The results show that Karect allows to map the largest number of reads, following by RECKONER, RACER, BLESS, and Blue.
Nevertheless, the absolute values are very close, especially for \emph{S.\ cerevisiae} dataset.

\begin{table}[!t]
\processtable{Results of mapping
\label{tab:results:mapping}}
{
\renewcommand{\tabcolsep}{0.5em}
\begin{tabular}{@{\extracolsep{0.45em}}lccc}
\toprule
&	\multicolumn{3}{c}{Organism}\\\cline{2-4}
Corrector & \emph{S.\ cerevisiae} & \emph{C.\ elegans} & \emph{M.\ acuminata} \\
\midrule
Karect				&	93.83		&\bf81.99	&\bf82.70\\	
RECKONER				&	93.76		&	81.58		&	82.46	\\	
RACER					&\bf93.99	&	81.79		&	82.04	\\	
BLESS					&	93.78		&	81.56		&	82.30	\\	
Blue					&	93.77		&	81.70		&	82.21	\\	
BFC					&	93.76		&	81.67		&	82.05	\\	
Musket				& 	93.69		&	81.54		&	82.31	\\	
Ace					&	93.78		&	81.52		&	82.10	\\	
Lighter				&	93.73		&	81.52		& 	82.30	\\	
Trowel				& 	92.92		&	81.56		&	82.26	\\	
Pollux				&	93.64		&	81.42		&	82.45	\\	
\it Without corr.	&\it93.65	&\it81.45	&\it82.04\\	
\botrule
\end{tabular}
}{
The given numbers are fractions (expressed in \%) of mapped reads.
The correctors are ordered according to average rank.
The best values are in bold
}
\end{table}

\begin{table*}[htp]
\processtable{Results of correction using simulated reads
\label{tab:results:simulated}}
{
\renewcommand{\tabcolsep}{0.5em}
\begin{tabular}{@{\extracolsep{0.45em}}lcrrrrrc|lcrrrrr}
\toprule
Corrector && Sensitivity	& Precision	& Gain	& RAM [GB]	& Time [s]	&& Corrector && Sensitivity	& Precision	& Gain	& RAM	[GB]	& Time [s]	\\
\midrule
\multicolumn{15}{c}{\bf \emph{S.\ cerevisiae}, coverage 30$\times$}\\
&&	\multicolumn{5}{c}{read length 100\,bp, $p = 2.0\%$}&&	&&	\multicolumn{5}{c}{read length 100\,bp, $p = 5.5\%$}\\\cline{3-7}\cline{11-15}
RECKONER	&&	\bf99.077	&	99.972	&	\bf99.049	&	0.536	&	83		&& RECKONER	&&	\bf95.718	&	99.974	&	\bf95.693	&	0.573	&	88		\\
BLESS	&&	98.069	&	99.991	&	98.060	&	0.835	&	119		&& BLESS	&&	93.525	&	99.990	&	93.516	&	0.927	&	109		\\
BFC	&&	77.030	&	99.996	&	77.026	&	1.148	&	83		&& BFC	&&	68.406	&	99.997	&	68.404	&	1.157	&	\bf87		\\
Lighter	&&	80.470	&	99.923	&	80.407	&	\bf0.065	&	94		&& Lighter	&&	70.303	&	99.930	&	70.254	&	\bf0.067	&	97		\\
RACER	&&	80.819	&	99.946	&	80.775	&	1.827	&	184		&& Trowel	&&	71.648	&	99.975	&	71.631	&	6.674	&	140		\\
Musket	&&	60.356	&	\bf100.000	&	60.356	&	0.303	&	187		&& RACER	&&	70.766	&	99.950	&	70.731	&	1.827	&	179		\\
Trowel	&&	74.831	&	99.990	&	74.823	&	4.205	&	\bf64		&& Musket	&&	51.886	&	\bf100.000	&	51.886	&	0.304	&	245		\\
Karect	&&	94.608	&	99.932	&	94.544	&	7.005	&	168		&& Ace	&&	91.672	&	99.921	&	91.600	&	2.701	&	2109		\\
Ace	&&	96.555	&	99.904	&	96.462	&	2.164	&	1890		&& Karect	&&	85.665	&	99.940	&	85.614	&	7.005	&	472		\\
Blue	&&	72.195	&	99.316	&	71.698	&	1.855	&	582		&& Blue	&&	61.544	&	99.519	&	61.246	&	2.872	&	640		\\
Pollux	&&	42.993	&	90.728	&	38.599	&	4.013	&	949		&& Pollux	&&	36.297	&	91.789	&	33.050	&	4.558	&	1060		\\
\midrule
\multicolumn{15}{c}{\bf \emph{C.\ elegans}, coverage 20$\times$}\\
&&	\multicolumn{5}{c}{read length 150\,bp, $p = 1.9\%$}&&	&&	\multicolumn{5}{c}{read length 151\,bp, $p = 4.2\%$}\\\cline{3-7}\cline{11-15}
BLESS	&&	95.776	&	99.997	&	95.774	&	1.993	&	579		&& RECKONER	&&	\bf96.757	&	99.992	&	\bf96.749	&	1.619	&	536		\\
RECKONER	&&	\bf98.137	&	99.968	&	\bf98.105	&	1.864	&	518		&& BLESS	&&	94.240	&	\bf100.000	&	94.240	&	1.940	&	697		\\
Lighter	&&	82.244	&	99.994	&	82.239	&	\bf0.438	&	508		&& Lighter	&&	63.117	&	99.994	&	63.113	&	\bf0.438	&	498		\\
BFC	&&	82.160	&	\bf99.999	&	82.159	&	3.702	&	409		&& BFC	&&	61.761	&	99.999	&	61.760	&	3.705	&	501		\\
Musket	&&	74.155	&	99.999	&	74.154	&	0.950	&	1294		&& Trowel	&&	76.982	&	99.778	&	76.810	&	15.364	&	\bf342		\\
Ace	&&	89.606	&	99.995	&	89.601	&	13.172	&	12787		&& Musket	&&	37.152	&	99.998	&	37.151	&	0.963	&	1915		\\
Karect	&&	93.654	&	99.978	&	93.634	&	39.443	&	4265		&& Karect	&&	96.203	&	99.985	&	96.188	&	39.471	&	3026		\\
RACER	&&	76.433	&	99.987	&	76.424	&	13.619	&	1047		&& Ace	&&	85.085	&	99.994	&	85.080	&	17.276	&	14389		\\
Trowel	&&	67.257	&	99.838	&	67.148	&	15.737	&	\bf143		&& RACER	&&	33.529	&	99.983	&	33.523	&	13.618	&	1189		\\
Blue	&&	42.958	&	97.513	&	41.863	&	11.543	&	5315		&& Blue	&&	43.745	&	98.670	&	43.156	&	11.702	&	7612		\\
Pollux	&&	41.097	&	75.587	&	27.823	&	15.262	&	22428		&& Pollux	&&	30.039	&	80.880	&	22.938	&	30.657	&	23445		\\
\midrule
\multicolumn{15}{c}{\bf \emph{M.\ acuminata}, coverage 30$\times$}\\
&&	\multicolumn{5}{c}{read length 150\,bp, $p = 1.9\%$}&&	&&	\multicolumn{5}{c}{read length 151\,bp, $p = 4.2\%$}\\\cline{3-7}\cline{11-15}
BLESS	&&	\bf93.574	&	99.573	&	\bf93.172	&	3.717	&	3051		&& RECKONER	&&	\bf92.820	&	99.198	&	\bf92.070	&	3.725	&	2285		\\
BFC	&&	79.321	&	99.873	&	79.221	&	13.186	&	\bf1622		&& BLESS	&&	88.800	&	99.758	&	88.584	&	3.727	&	3310		\\
RECKONER	&&	93.328	&	98.616	&	92.018	&	3.723	&	2021		&& BFC	&&	58.549	&	99.910	&	58.496	&	13.184	&	1835		\\
Lighter	&&	78.215	&	98.964	&	77.396	&	\bf1.336	&	1977		&& Lighter	&&	57.952	&	99.157	&	57.460	&	\bf1.336	&	2023		\\
Musket	&&	71.614	&	\bf100.000	&	71.614	&	3.328	&	5175		&& Blue	&&	68.392	&	98.642	&	67.451	&	37.104	&	\bf1380		\\
Karect	&&	88.183	&	97.487	&	85.910	&	103.137	&	5627		&& Musket	&&	35.366	&	\bf99.999	&	35.366	&	3.386	&	8627		\\
Blue	&&	67.139	&	97.469	&	65.395	&	35.010	&	1669		&& Karect	&&	89.751	&	98.502	&	88.387	&	103.421	&	12352		\\
RACER	&&	72.467	&	97.436	&	70.560	&	45.066	&	5056		&& Trowel	&&	46.441	&	97.716	&	45.355	&	57.607	&	2440		\\
Trowel	&&	48.584	&	96.107	&	46.616	&	57.547	&	1936		&& RACER	&&	40.113	&	97.305	&	39.002	&	45.060	&	4754		\\
Pollux	&&	\mcOT	&& Pollux	&&	\mcOT		\\
Ace	&&	\mcOT		&& Ace	&&	\mcOT		\\
\botrule
\end{tabular}
}{
The values $p$ are average probabilities of base error.
The correctors are ordered according to average rank.
The best values are in bold
}
\end{table*}

\subsection{Simulated data evaluation}
The most direct method of evaluation of correction algorithms is performing tests on reads generated \emph{in silico},
so we simulated genome sequencing by generating
a set of ideal reads and introducing errors to them.

The applied method of reads generation was proposed and used in \citep{KSS10} and \citep{LSS13}.
First of all, we needed to choose the expected coverage and read length.
Then, we excerpted fragments of the specified length from a full reference genome.
To introduce changes imitating sequencing errors we took real
quality indicators from FASTQ files with reads of length of reads
being generated. Details of datasets being a source of probabilities---patterns---are given
in the Supplementary Table~2. 
Patterns have been selected to represent
reads of lengths about $100 \text{bp}$ and $150 \text{bp}$ and two levels
of data quality.
(The reported base error probability was calculated as an average among the whole FASTQ file.)
Generation has been performed for three
differently sized organisms: \emph{S.~cerevisiae}, \emph{C.~elegans}, and \emph{M.~acuminata}
for different read coverages (20$\times$ and 30$\times$).

The parameters of algorithms, especially $k$-mer lengths, have been determined empirically by choosing
the best value for a specified algorithm (according to the preliminary experiments; results not shown).



Each corrected read was assigned to one of the following categories:
\begin{itemize}
\item $\tp$ (true positive)---before correction the read contained errors
and it was perfectly corrected,
\item $\tn$ (true negative)---before correction the read contained no errors
and it remained error-free after correction,
\item $\fp$ (false positive)---before correction the read contained no errors,
but the algorithm introduced at least one error to it,
\item $\fn$ (false negative)---before correction the read contained errors,
but the algorithm was not able to correct it properly, i.e., errors were not corrected,
were miscorrected, or the algorithm introduced new errors.
\end{itemize}

For comparison of correction accuracy we used three statistical measurements: $\sensitivity={|\tp| / (|\tp|+|\fn|)}$,
$\precision={|\tp| / (|\tp|+|\fp|)}$, $\gain={|\tp|-|\fp| / (|\tp|+|\fn|)}$.

The selection of results is given in Table \ref{tab:results:simulated}. Full results and plots
are given in Supplementary material. Results show that RECKONER is in all cases in a group of the three
best algorithms and in four cases it is the best one, what is the effect of the highest values of $\gain$ and
$\sensitivity$ in most cases and values of time and memory being close to best.

RECKONER achieves best results for data of poorer quality. In all cases RECKONER, BLESS, BFC, and Lighter are
ranked in the top 4 places.\looseness=-1

The important observation is Karect's memory requirements. Karect allows to specify 
the upper limit of memory consumption (we limited it to 120~GB), but choosing values
acceptable for a typical PC (e.g., 16 or 32~GB) causes considerable increase of computational time
and moderate decrease of quality.


\end{methods}
\section{Conclusion}
We have presented RECKONER, an efficient error corrector for the sequencing data.
It is based on the BLESS code.
Nevertheless, RECKONER implements several new ideas that allowed us to obtain highly competitive results when compared to the state-of-the-art algorithms.
In simulated-data experiments it was usually the best according to the gain and sensitivity measures.
RECKONER is also among the fastest and most memory frugal algorithms that allows to run it even on a commodity personal computer for quite large data.

In the real-data experiments, in which we evaluated both the quality of mapping of corrected reads and quality of \emph{de novo} assembly, it was at the second place, just after Karect.
Nevertheless, the memory requirements of Karect are much larger, which limits its applications to rather powerful computing servers.


\section*{Funding}
This work was supported by the Polish National Science Centre under the project DEC-2012/05/B/ST6/03148.
The work was performed using the
infrastructure supported by POIG.02.03.01-24-099/13 grant:
``GeCONiI--Upper Silesian Center for Computational Science and
Engineering''.


\clearpage
\pagestyle{empty}
\includegraphics{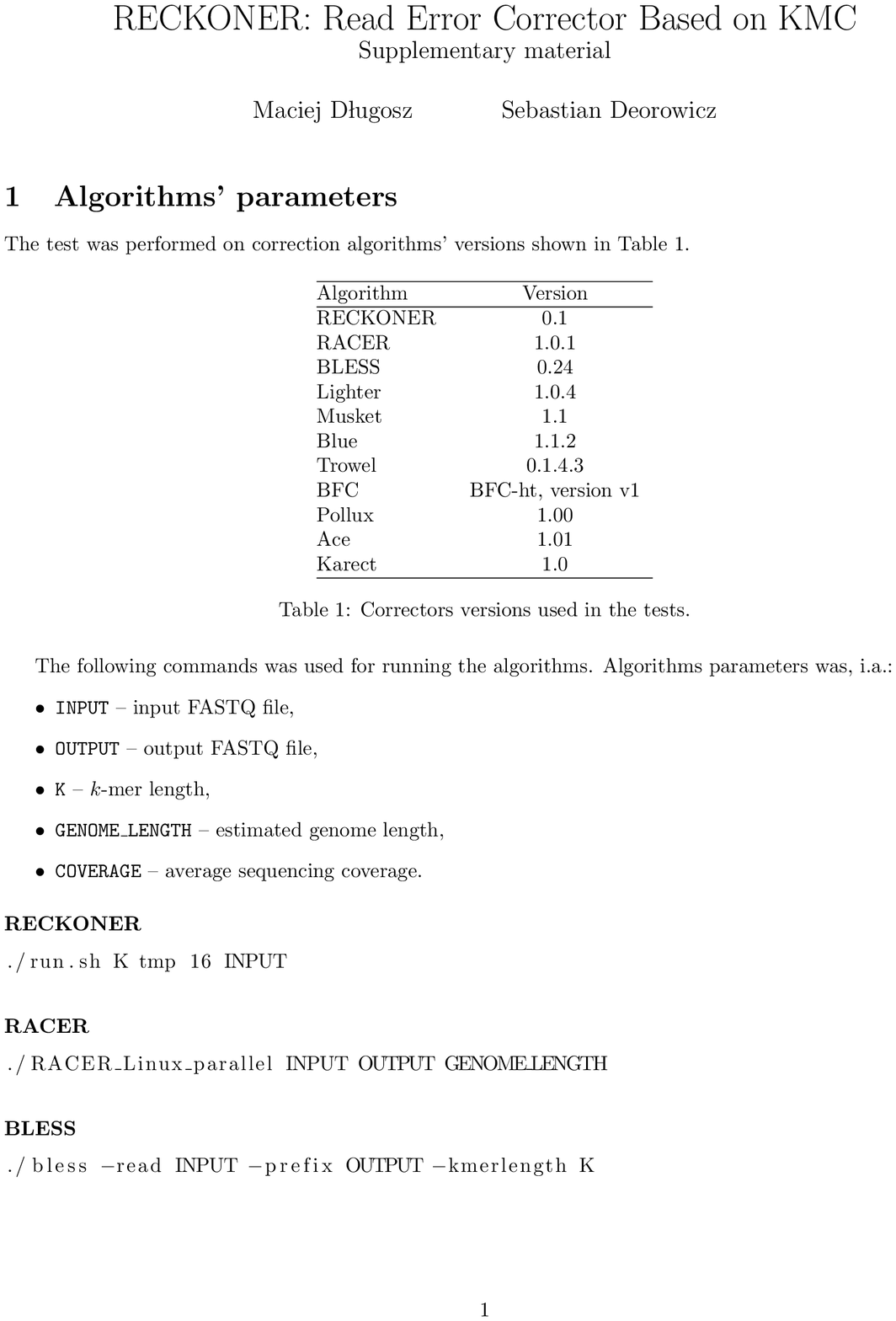}
\clearpage
\includegraphics{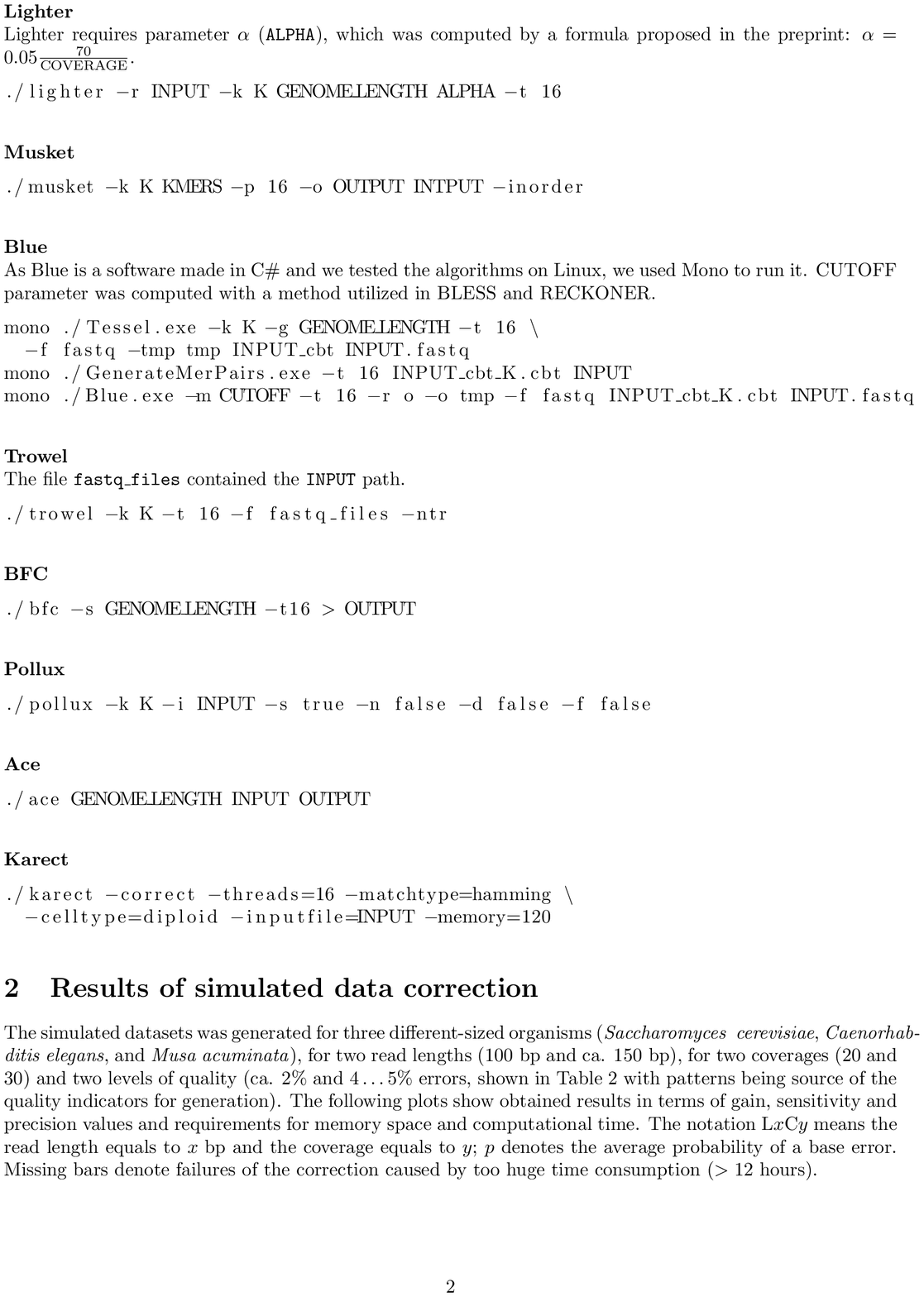}
\clearpage
\includegraphics{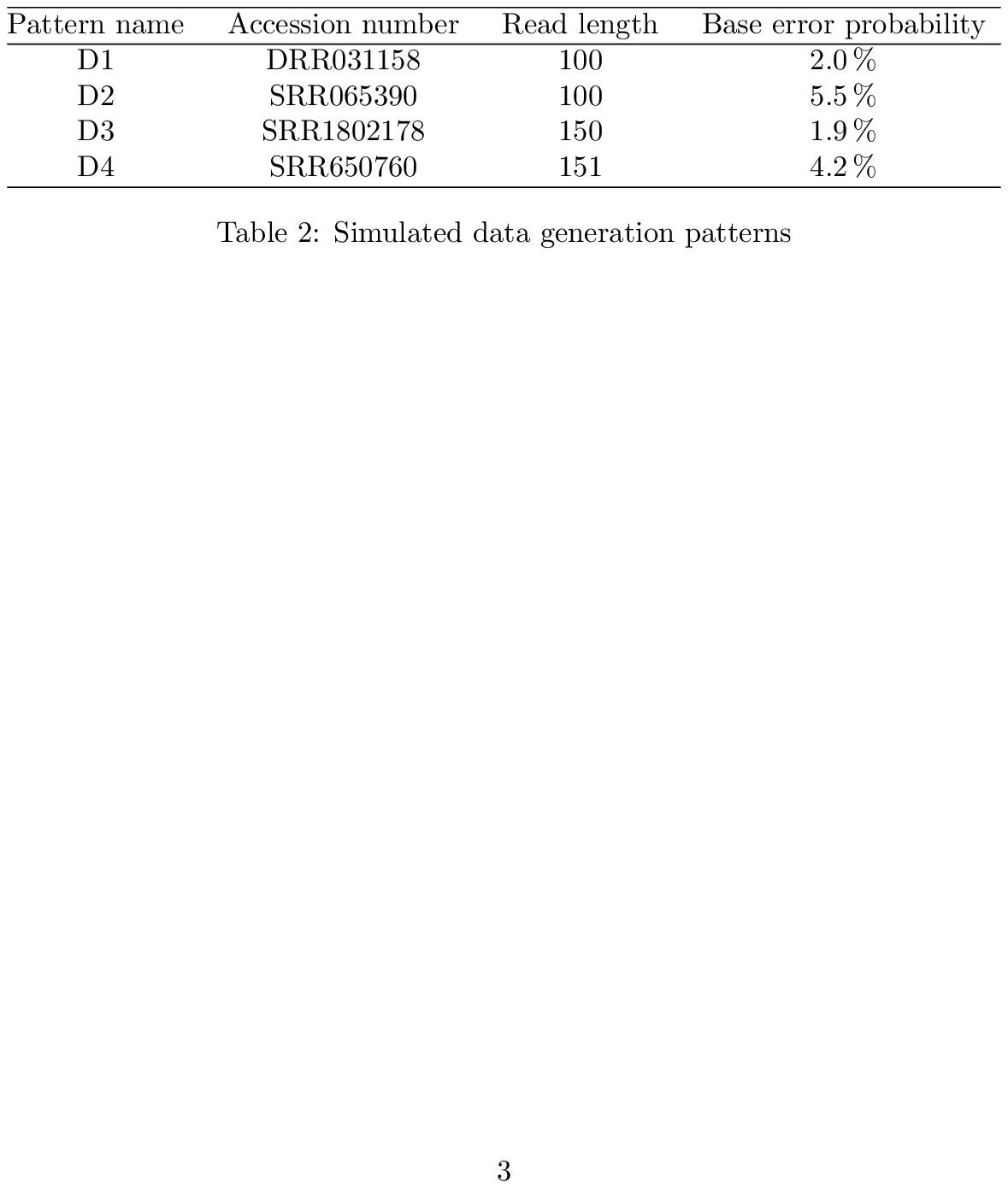}
\clearpage
\begin{figure}[p]
\includegraphics{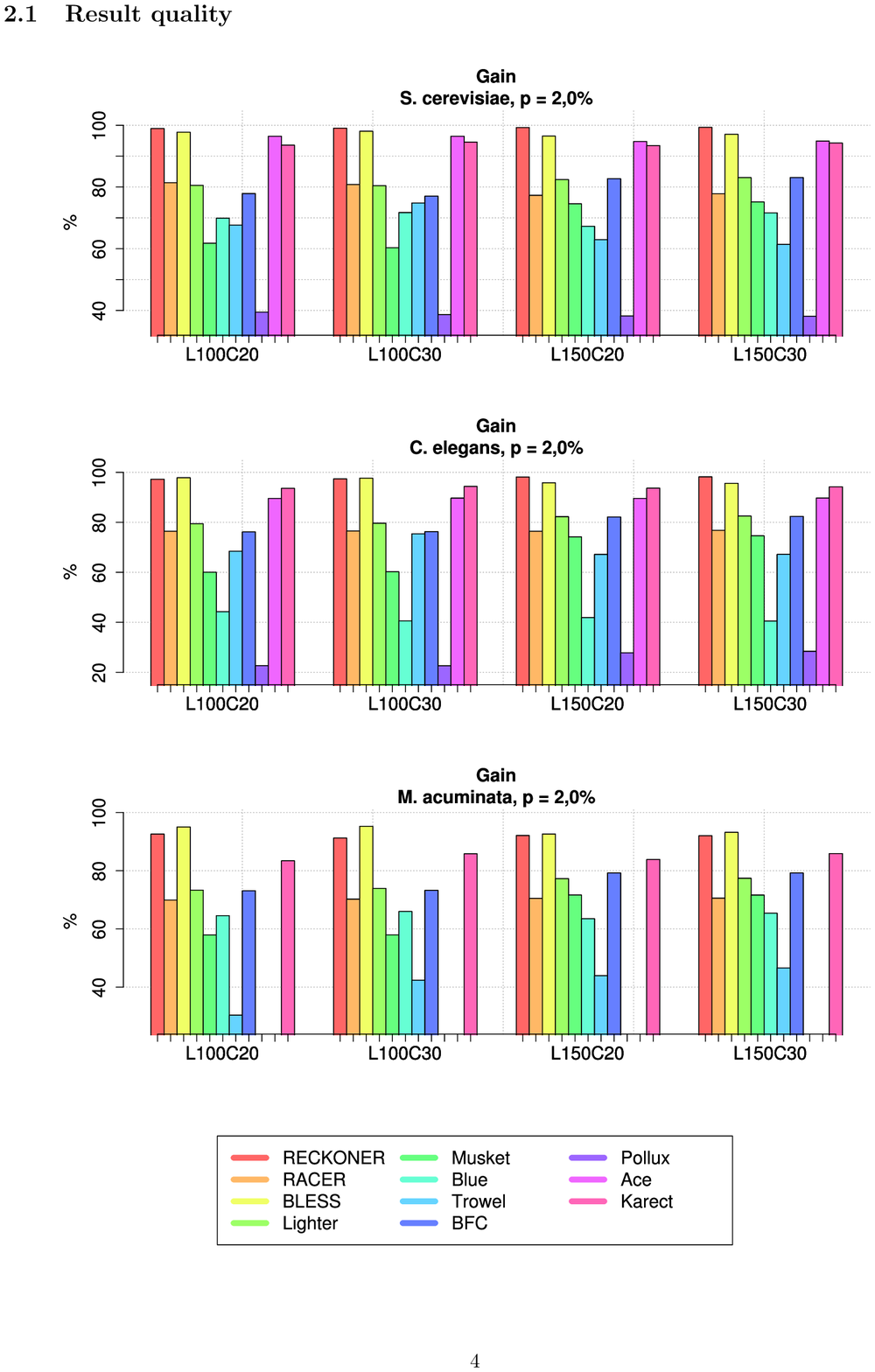}
\end{figure}
\clearpage
\begin{figure}[p]
\includegraphics{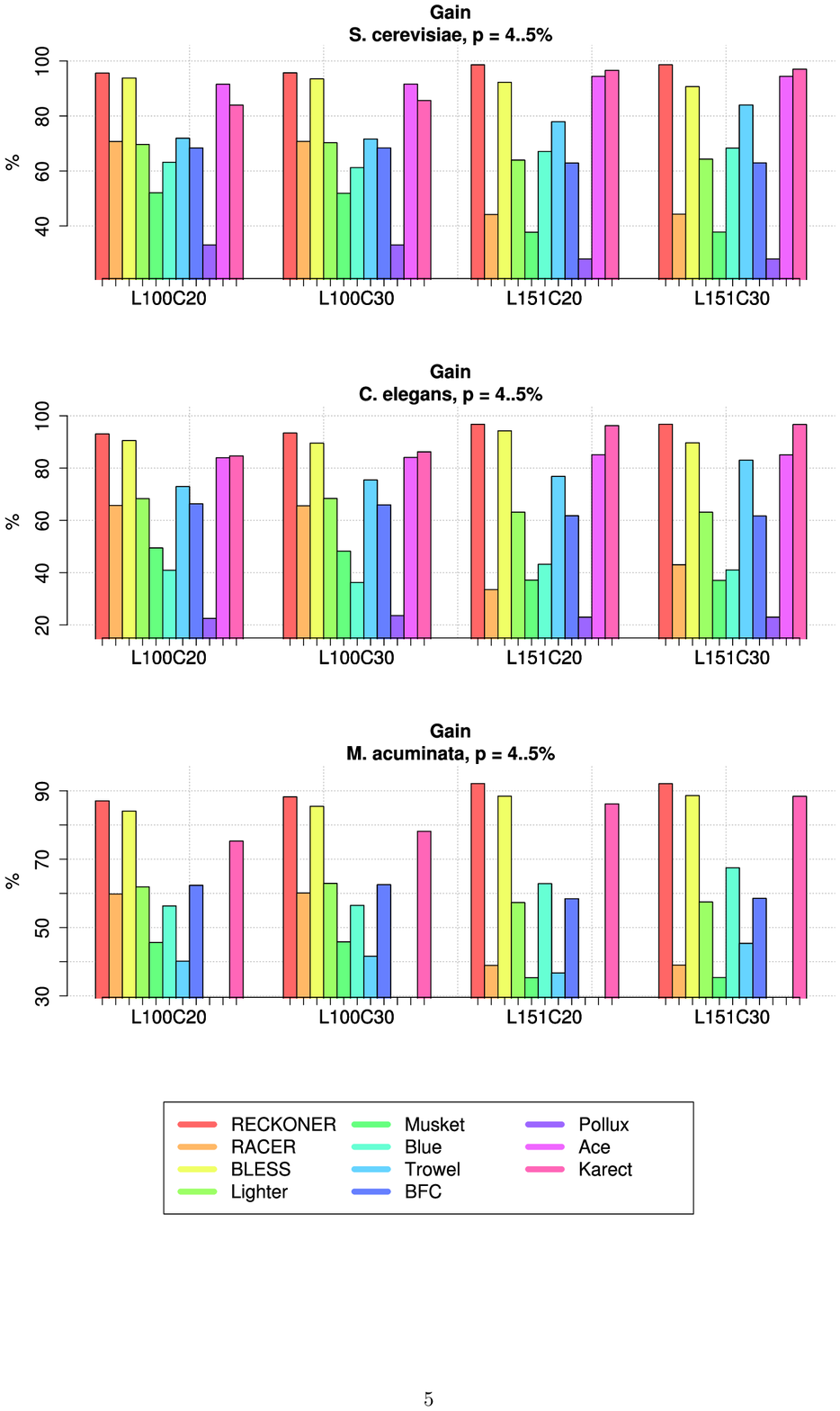}
\end{figure}
\clearpage
\begin{figure}[p]
\includegraphics{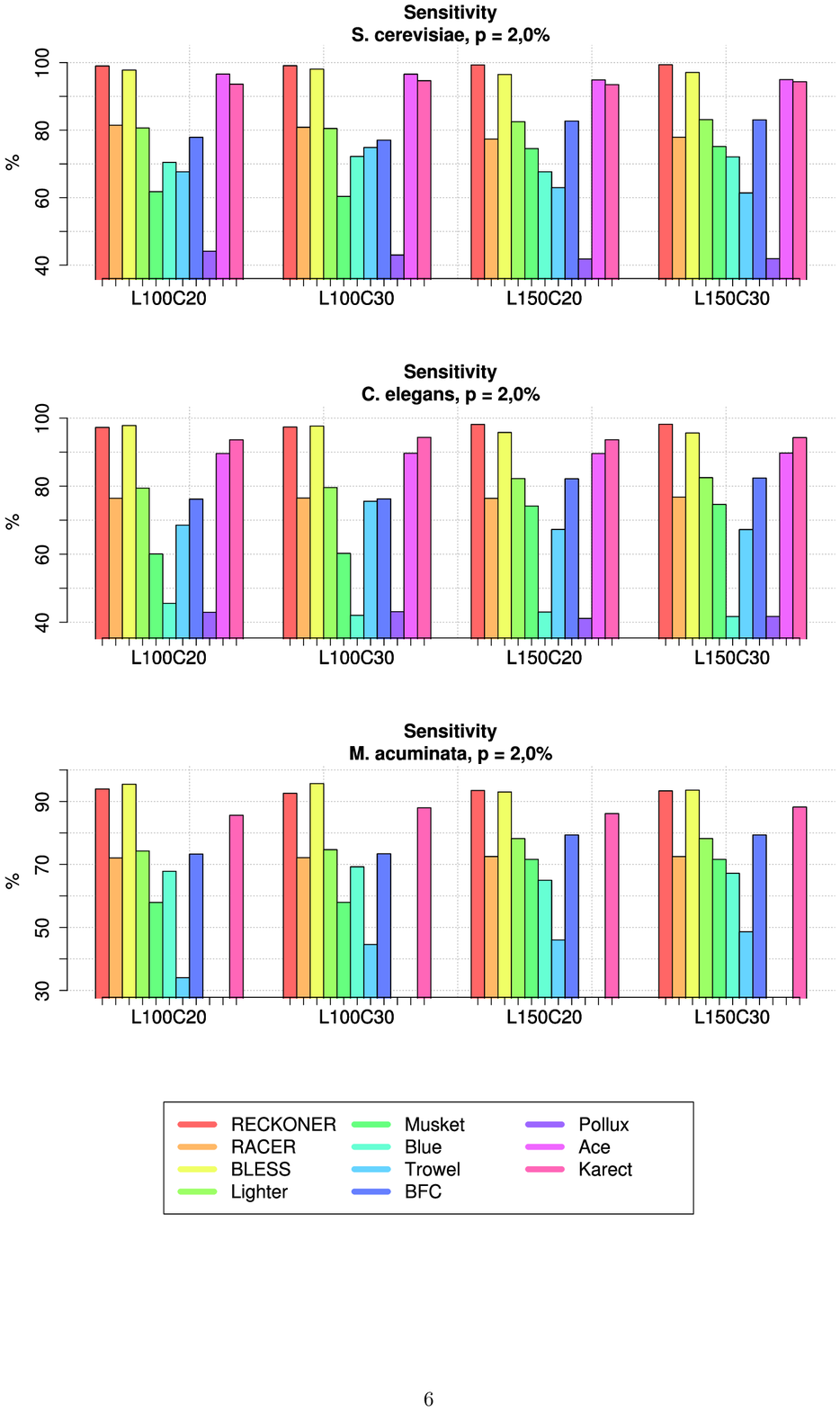}
\end{figure}
\clearpage
\begin{figure}[p]
\includegraphics{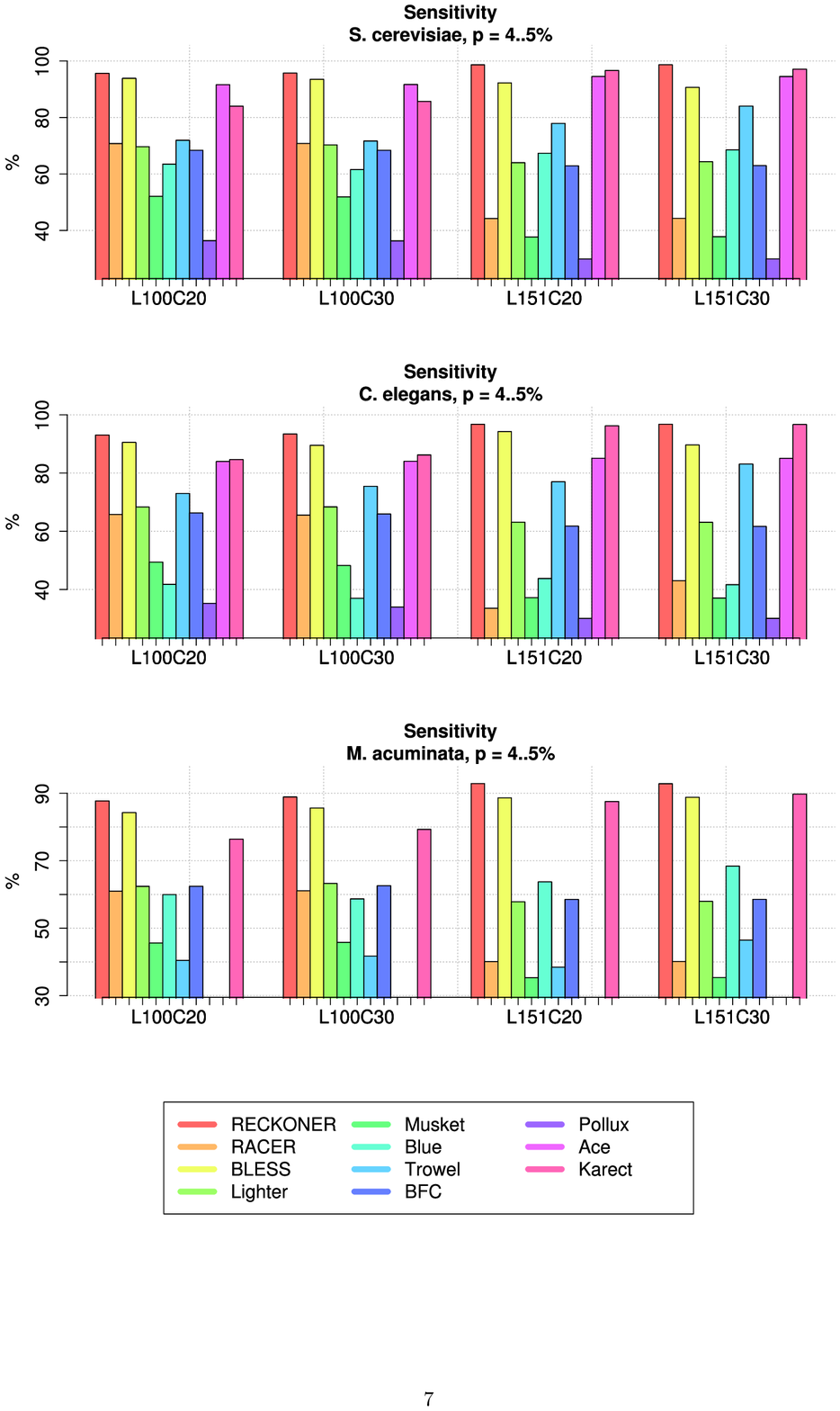}
\end{figure}
\clearpage
\begin{figure}[p]
\includegraphics{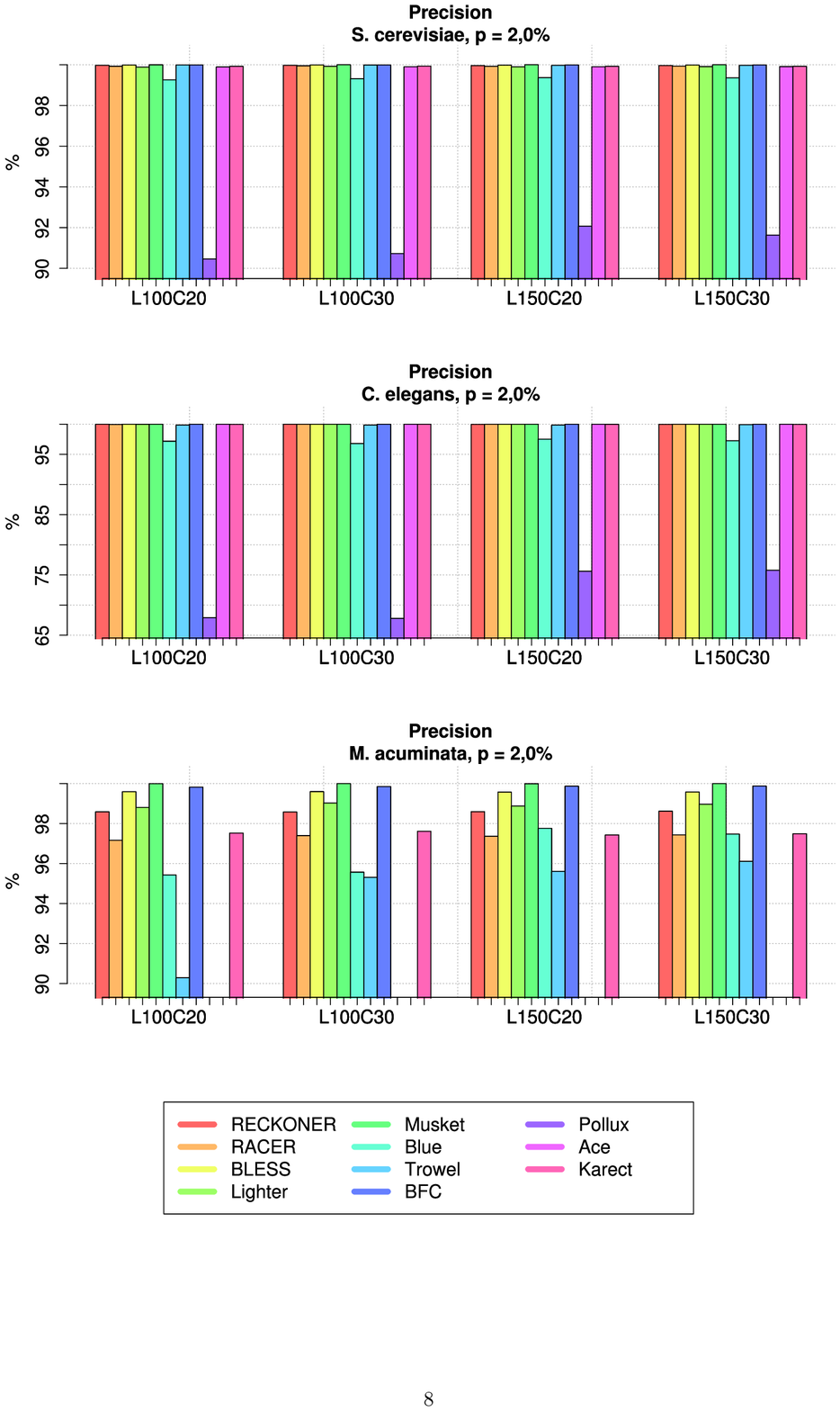}
\end{figure}
\clearpage
\begin{figure}[p]
\includegraphics{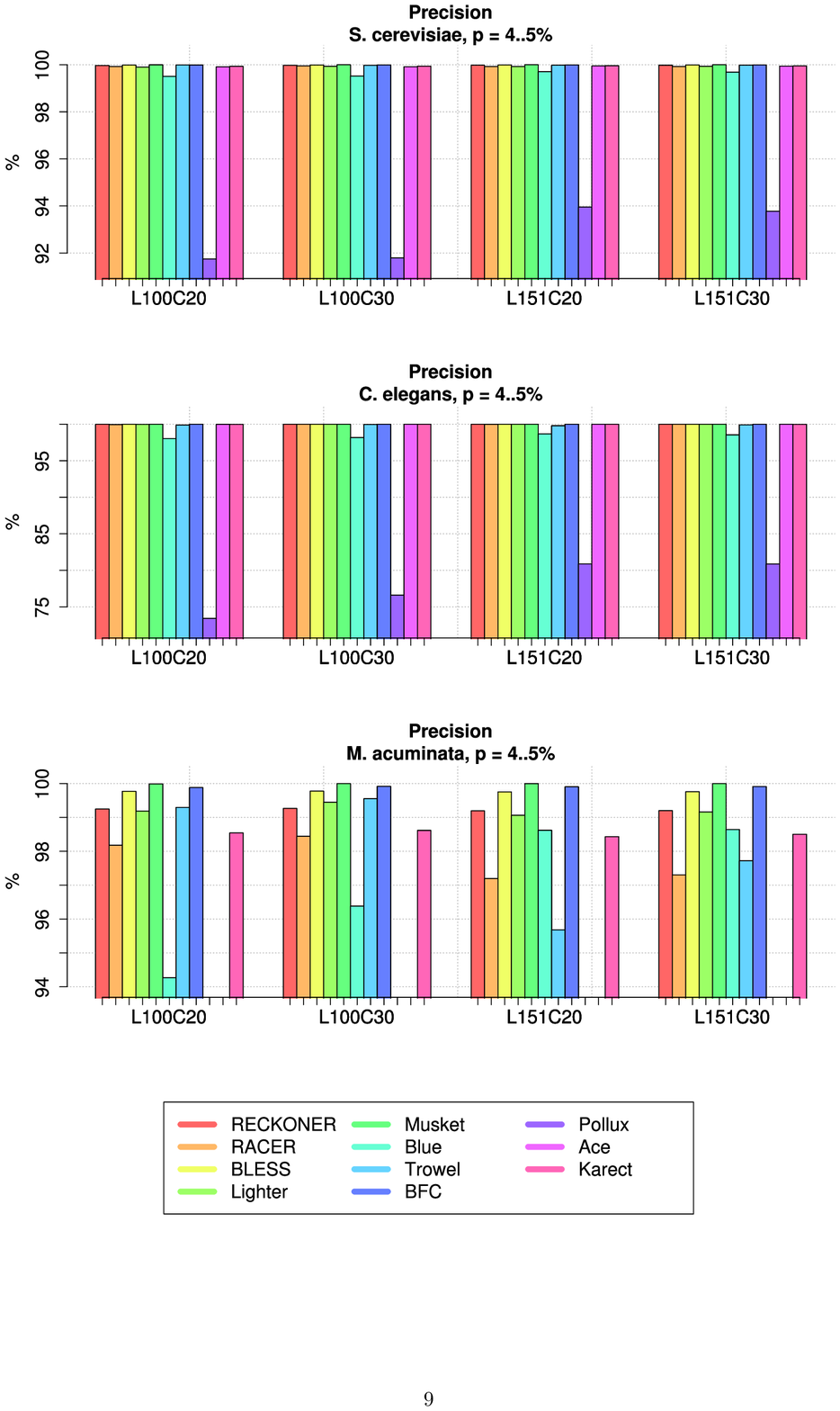}
\end{figure}
\clearpage
\begin{figure}[p]
\includegraphics{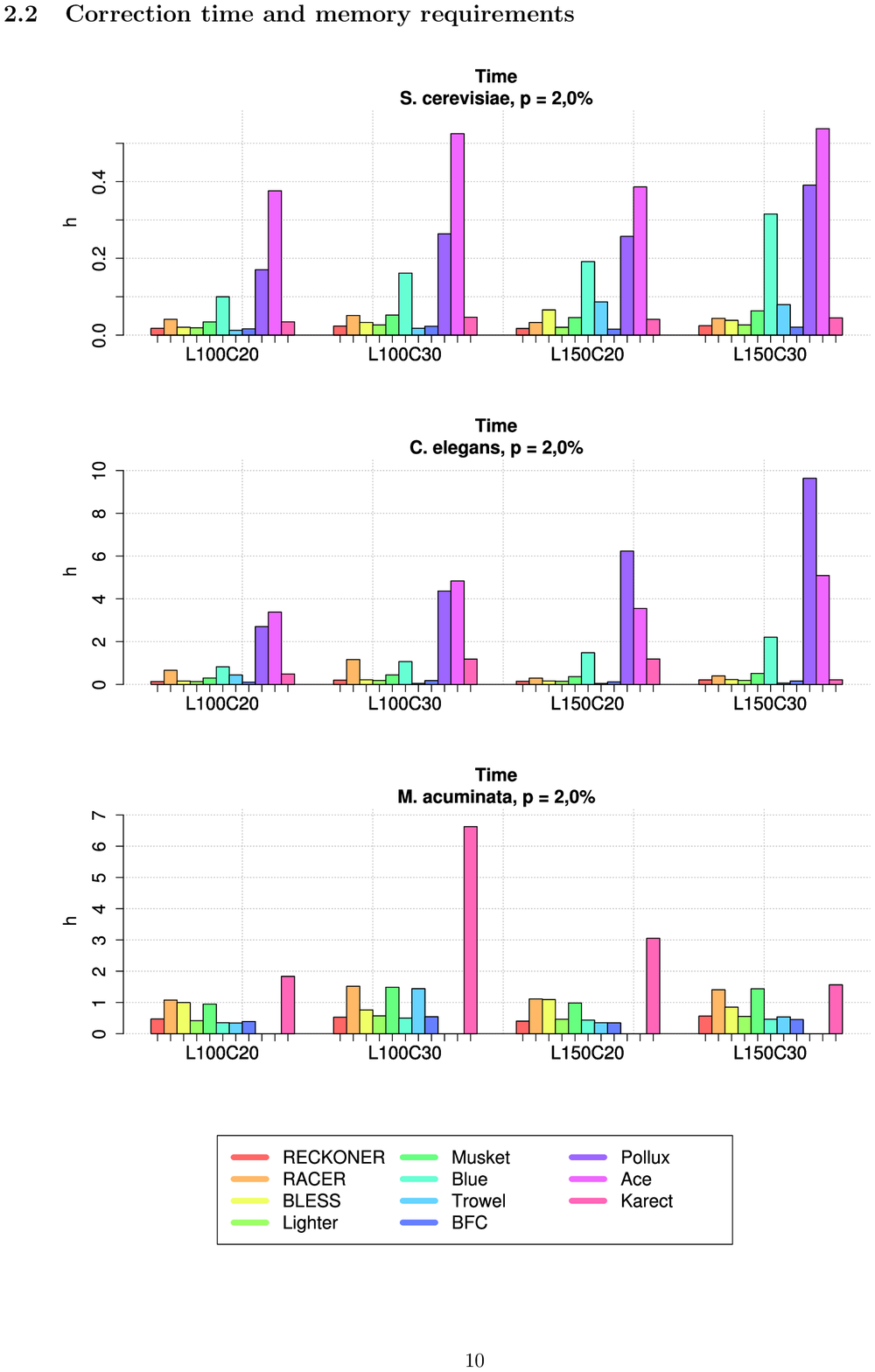}
\end{figure}
\clearpage
\includegraphics{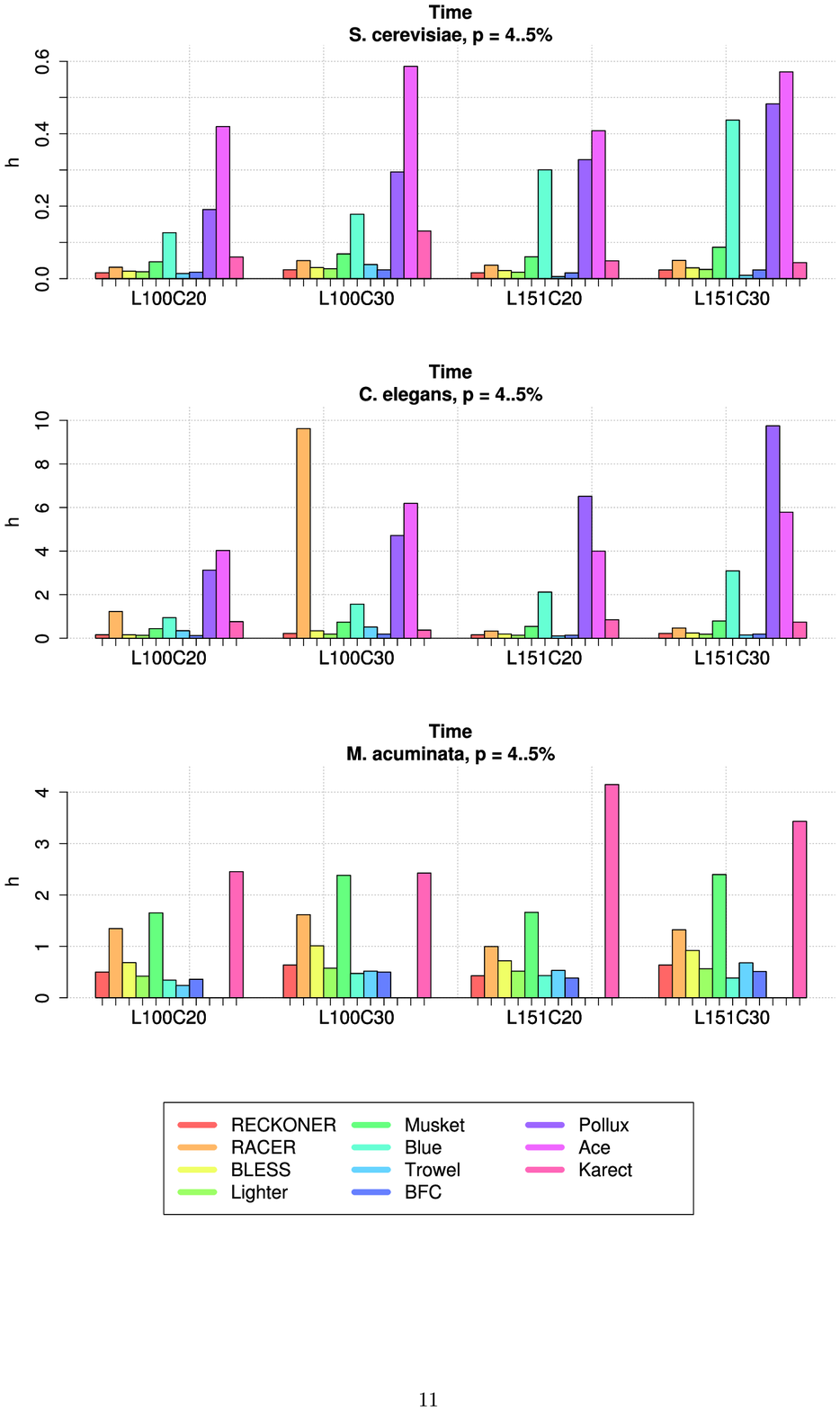}
\clearpage
\includegraphics{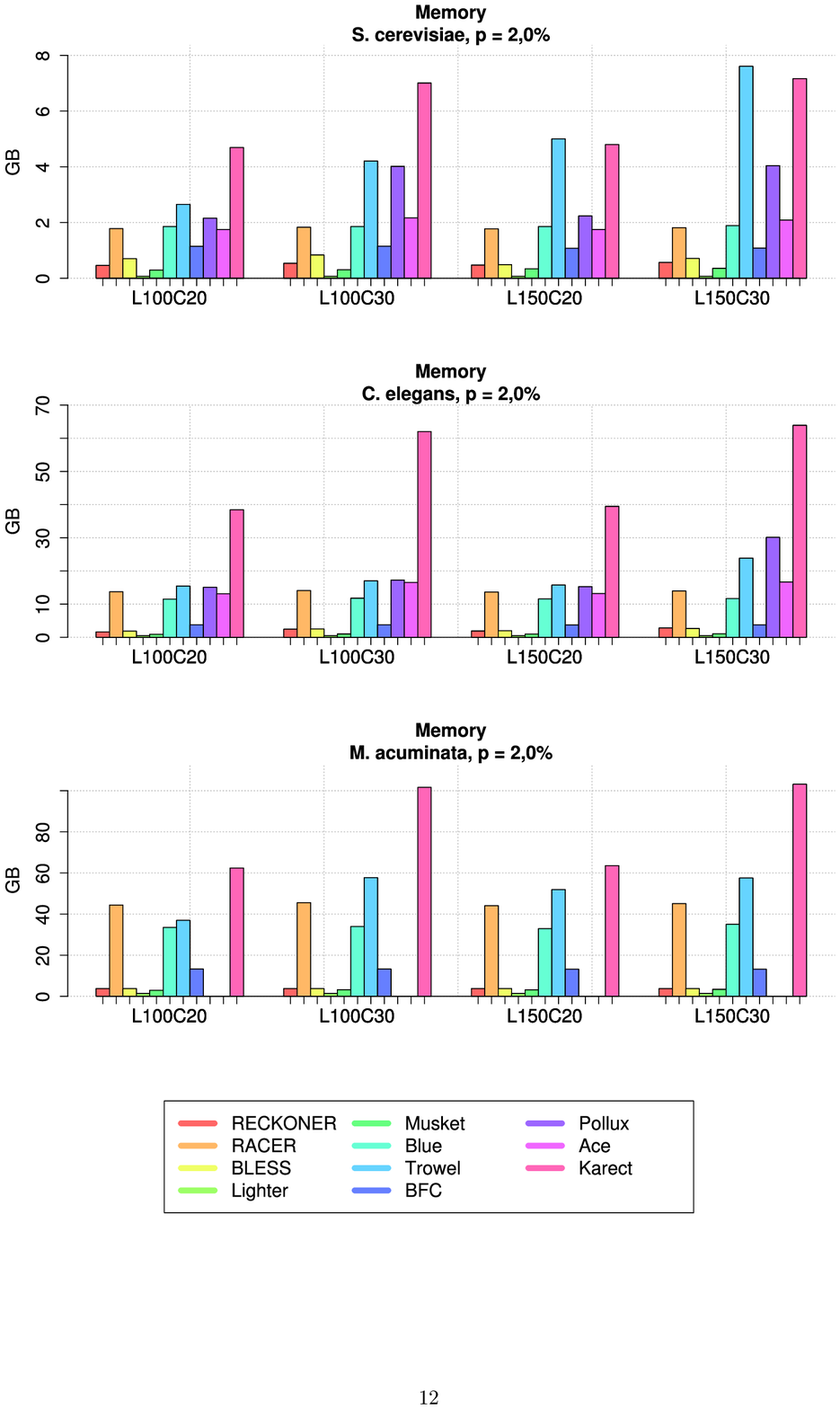}
\clearpage
\begin{figure}[p]
\includegraphics{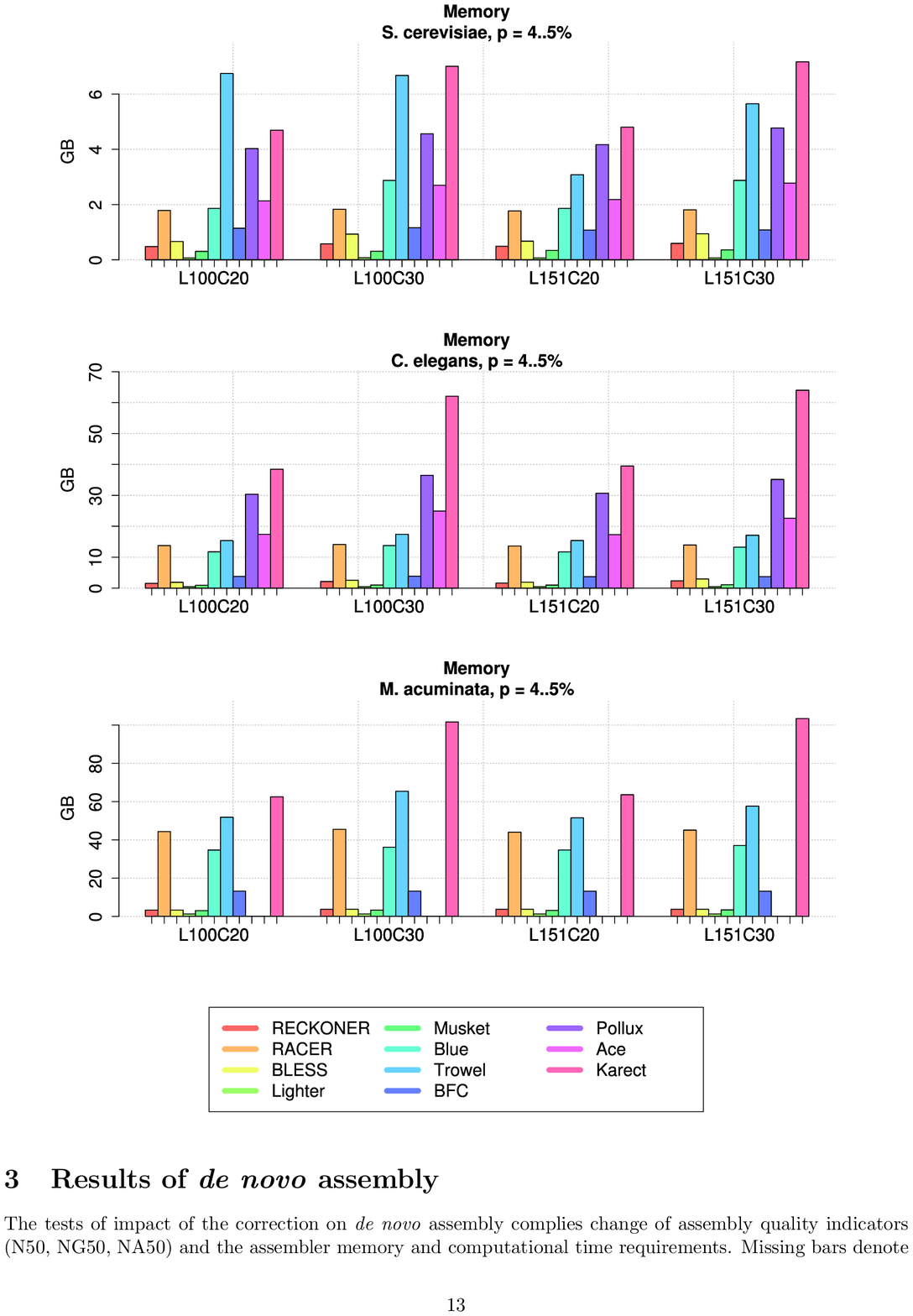}
\end{figure}
\clearpage
\includegraphics{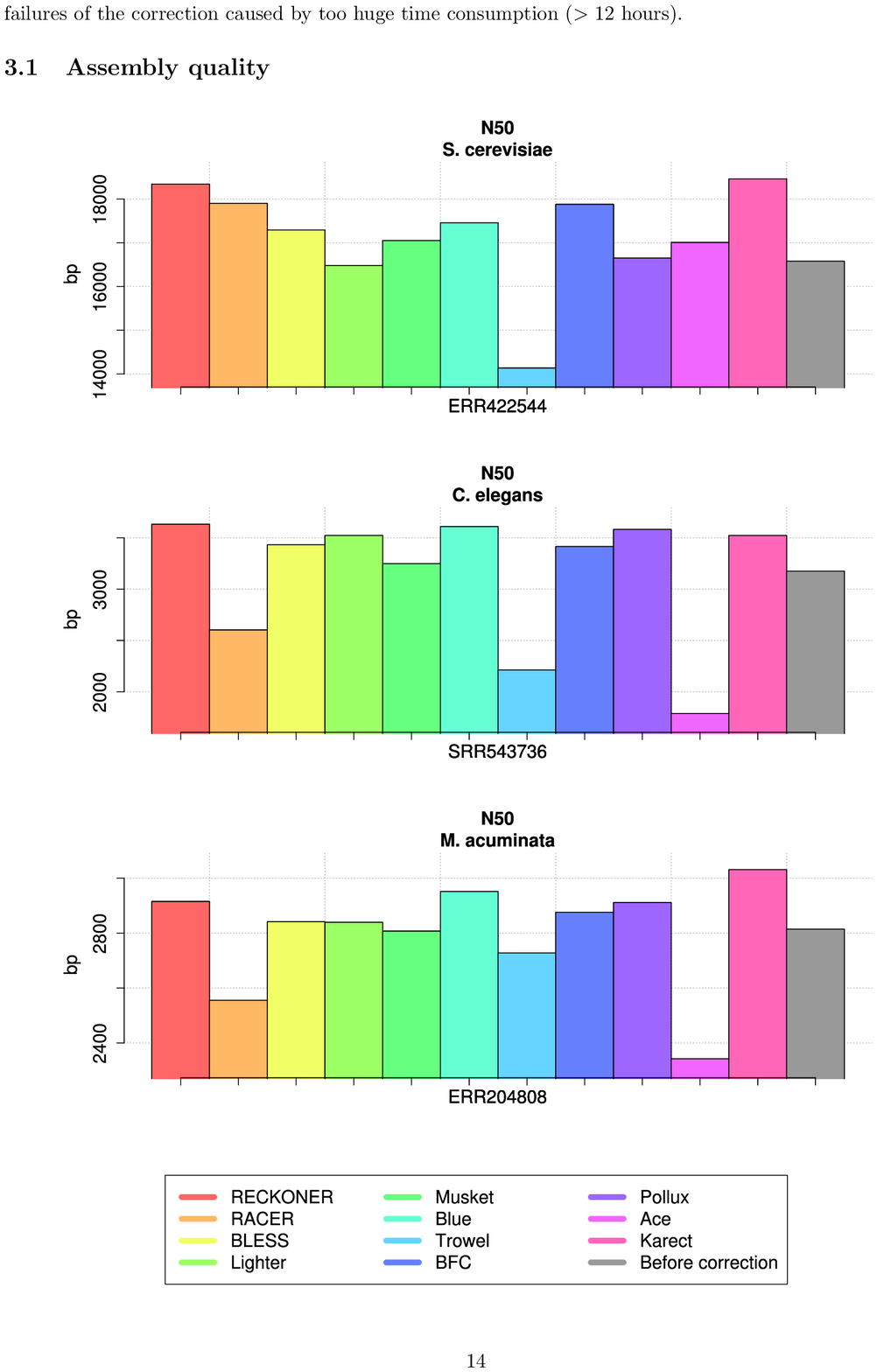}
\clearpage
\begin{figure}[p]
\includegraphics{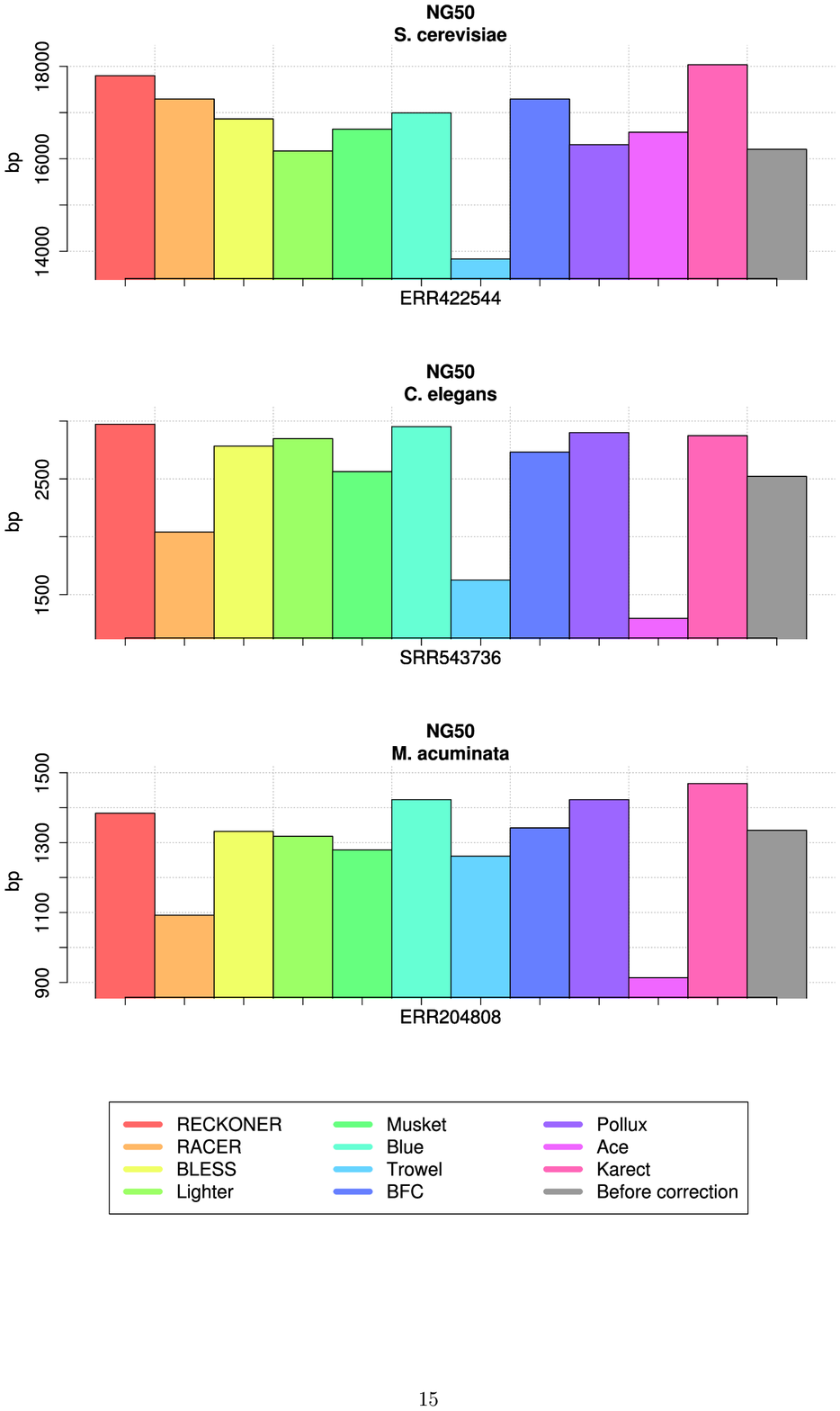}
\end{figure}
\clearpage
\begin{figure}[p]
\includegraphics{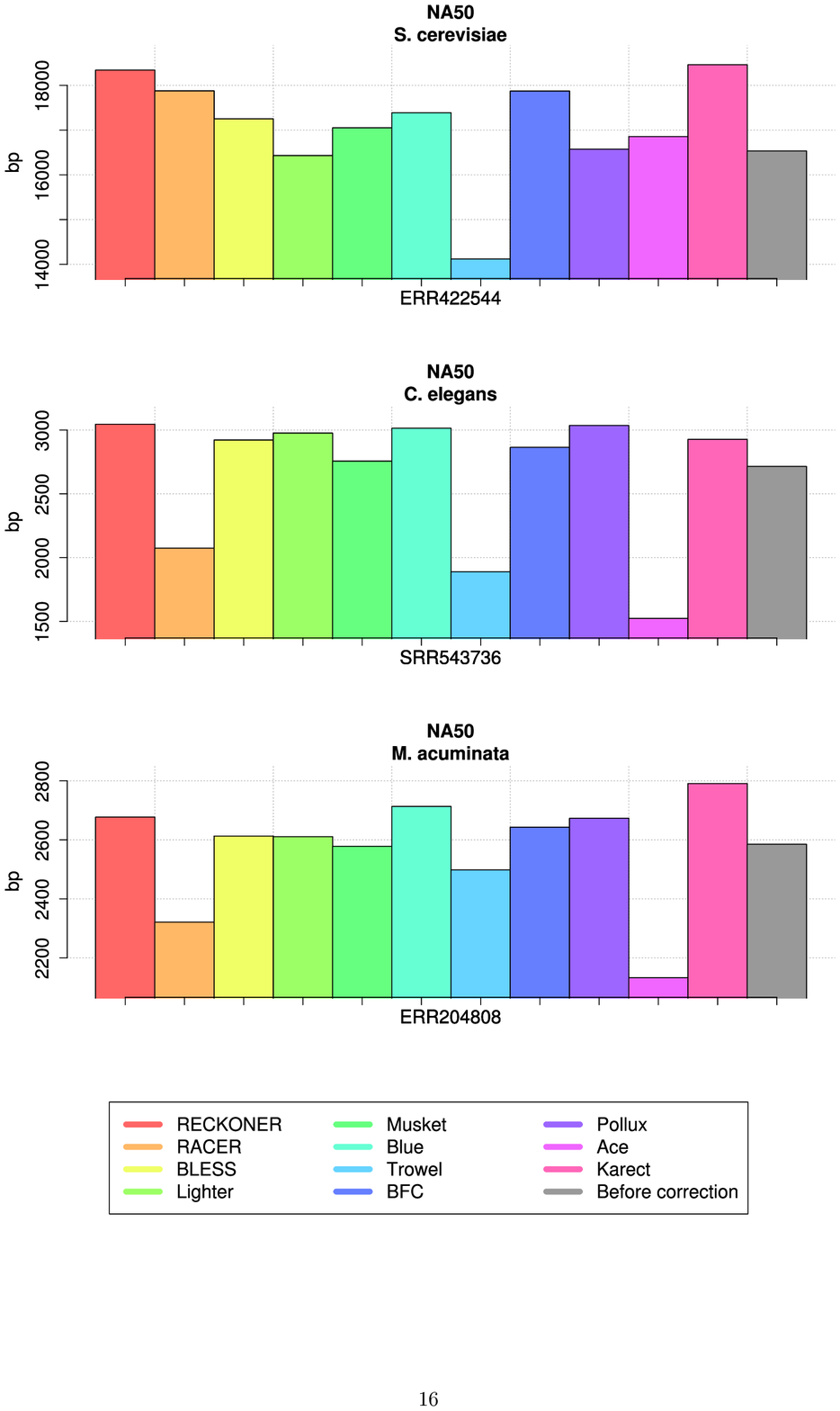}
\end{figure}
\clearpage
\begin{figure}[p]
\includegraphics{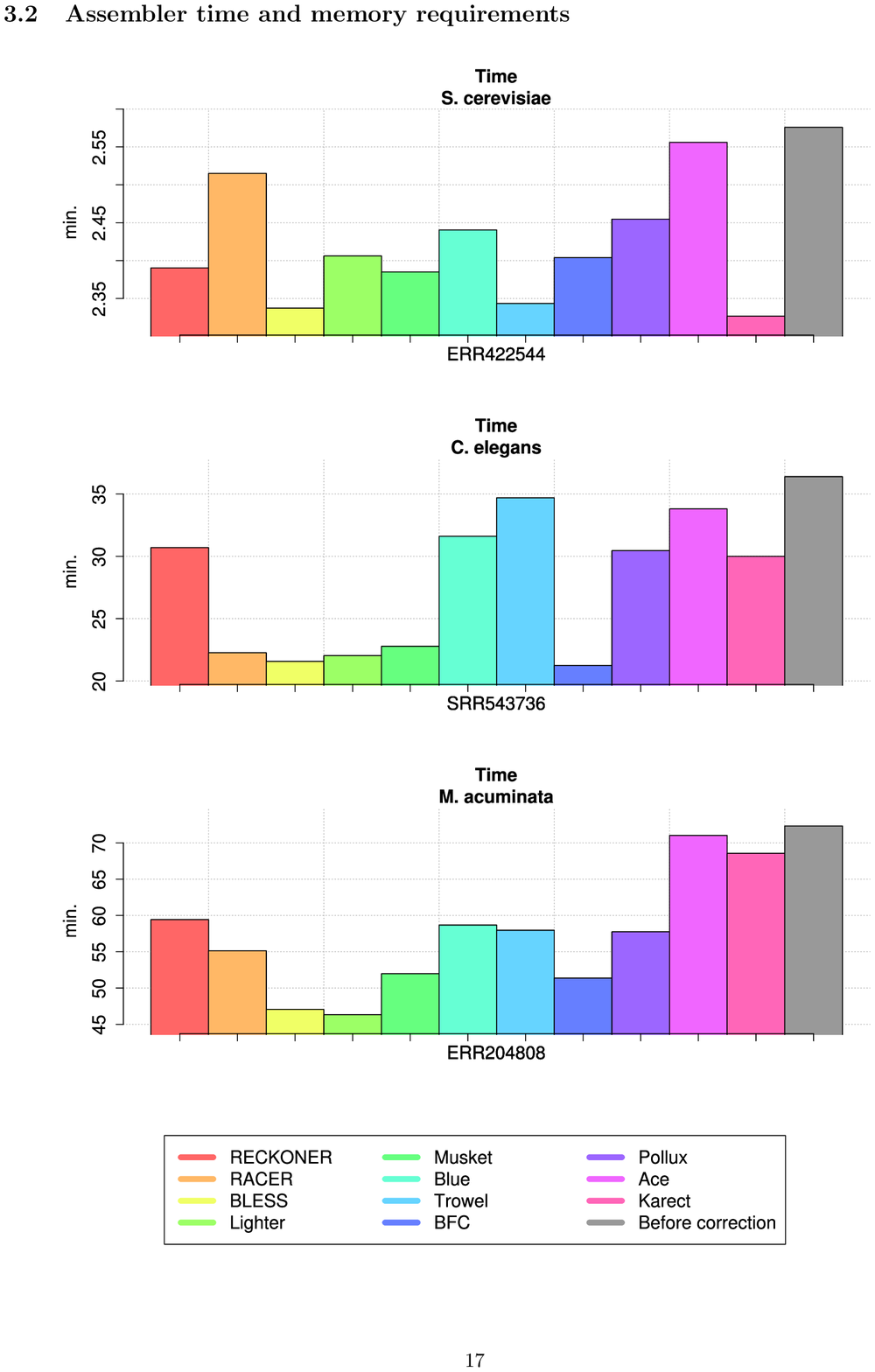}
\end{figure}
\clearpage
\begin{figure}[p]
\includegraphics{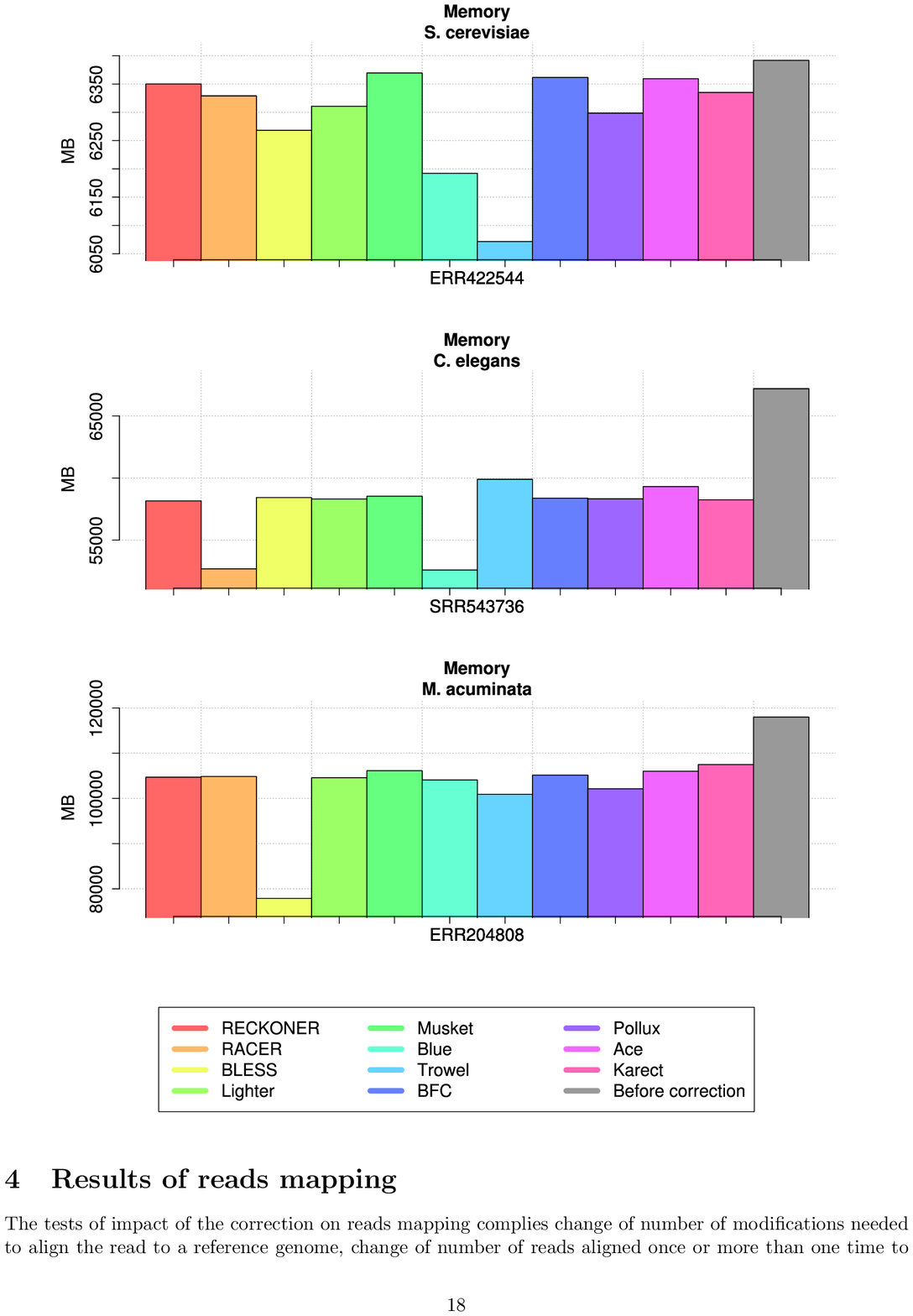}
\end{figure}
\clearpage
\begin{figure}[p]
\includegraphics{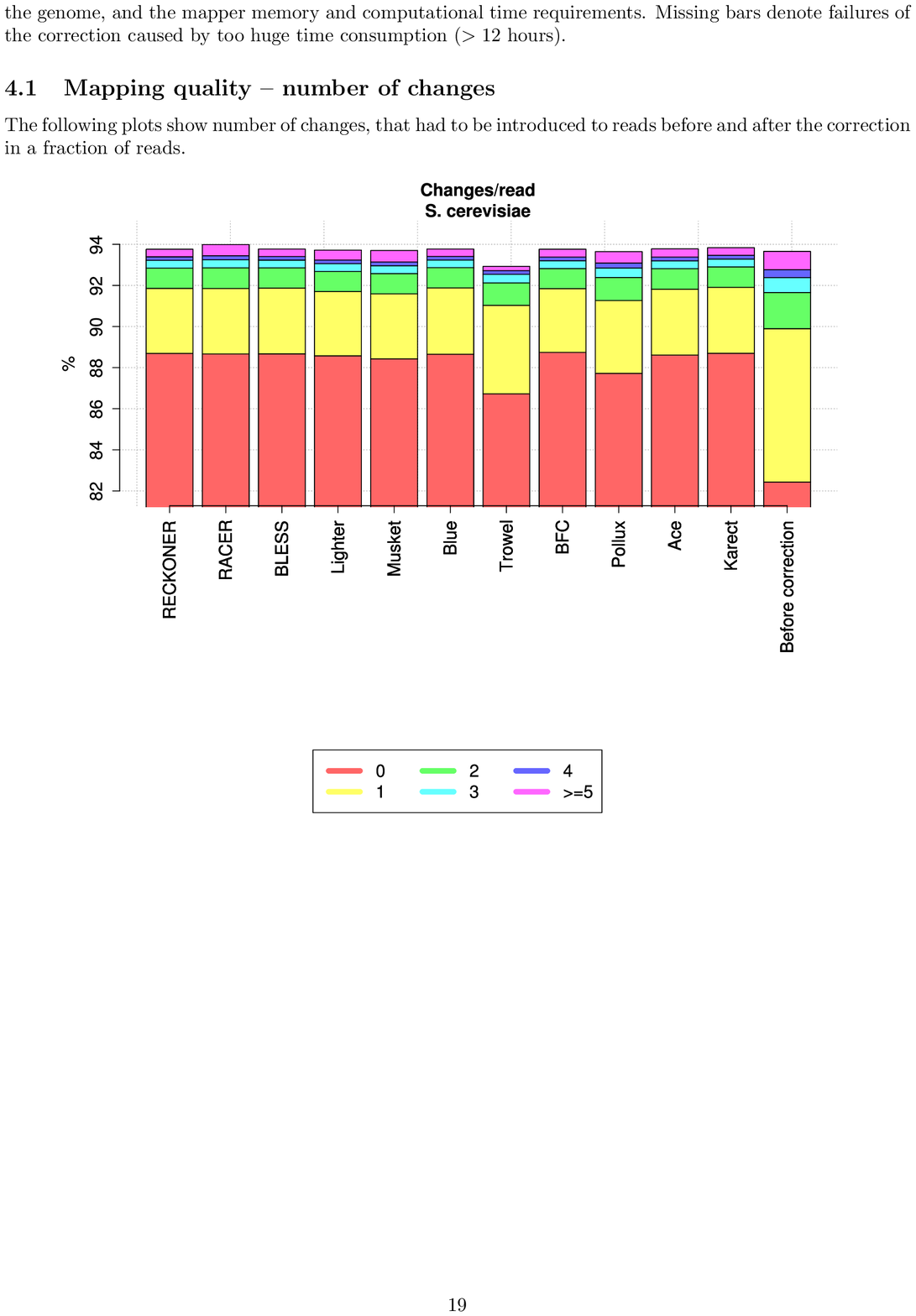}
\end{figure}
\clearpage
\begin{figure}[p]
\includegraphics{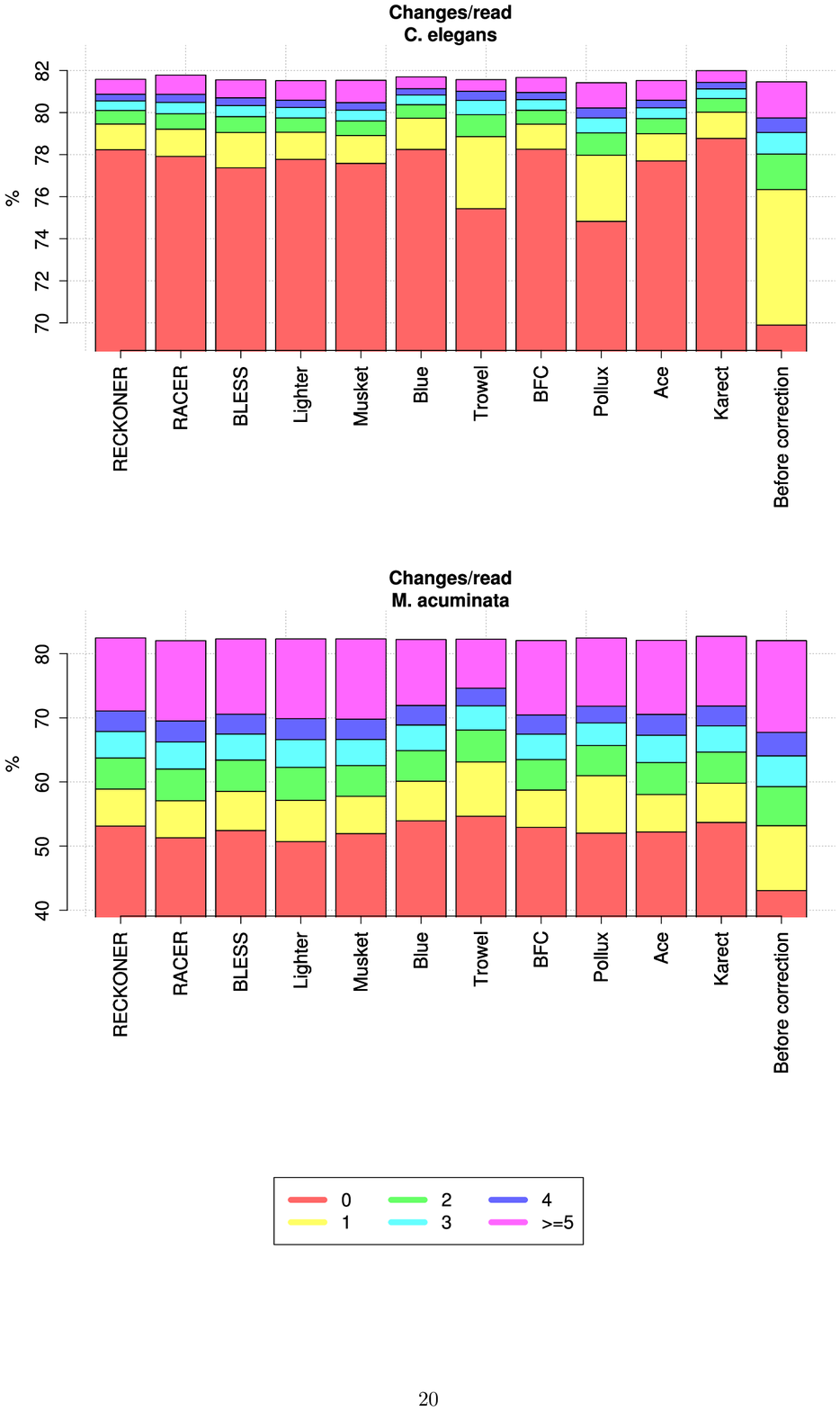}
\end{figure}
\clearpage
\begin{figure}[p]
\includegraphics{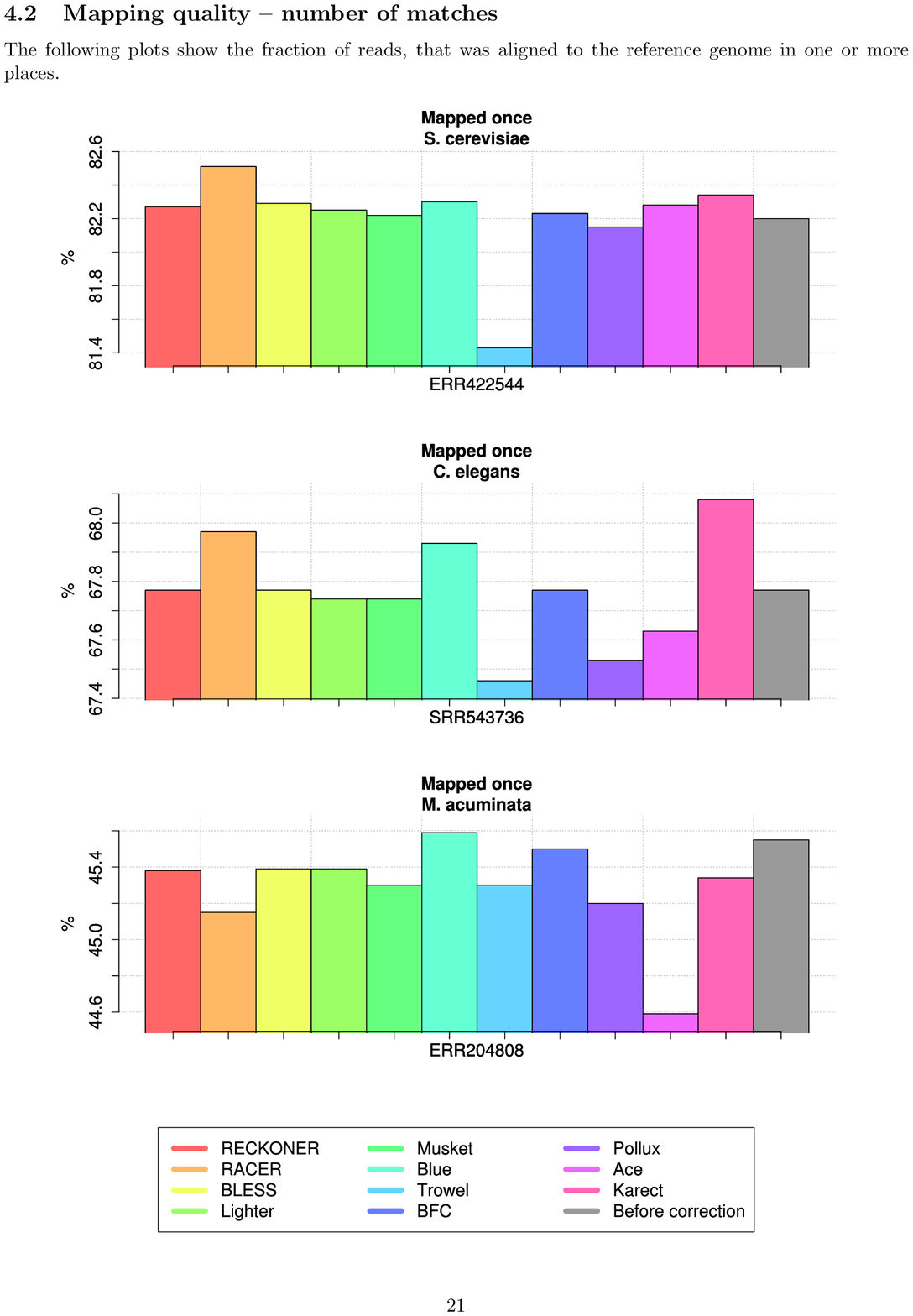}
\end{figure}
\clearpage
\begin{figure}[p]
\includegraphics{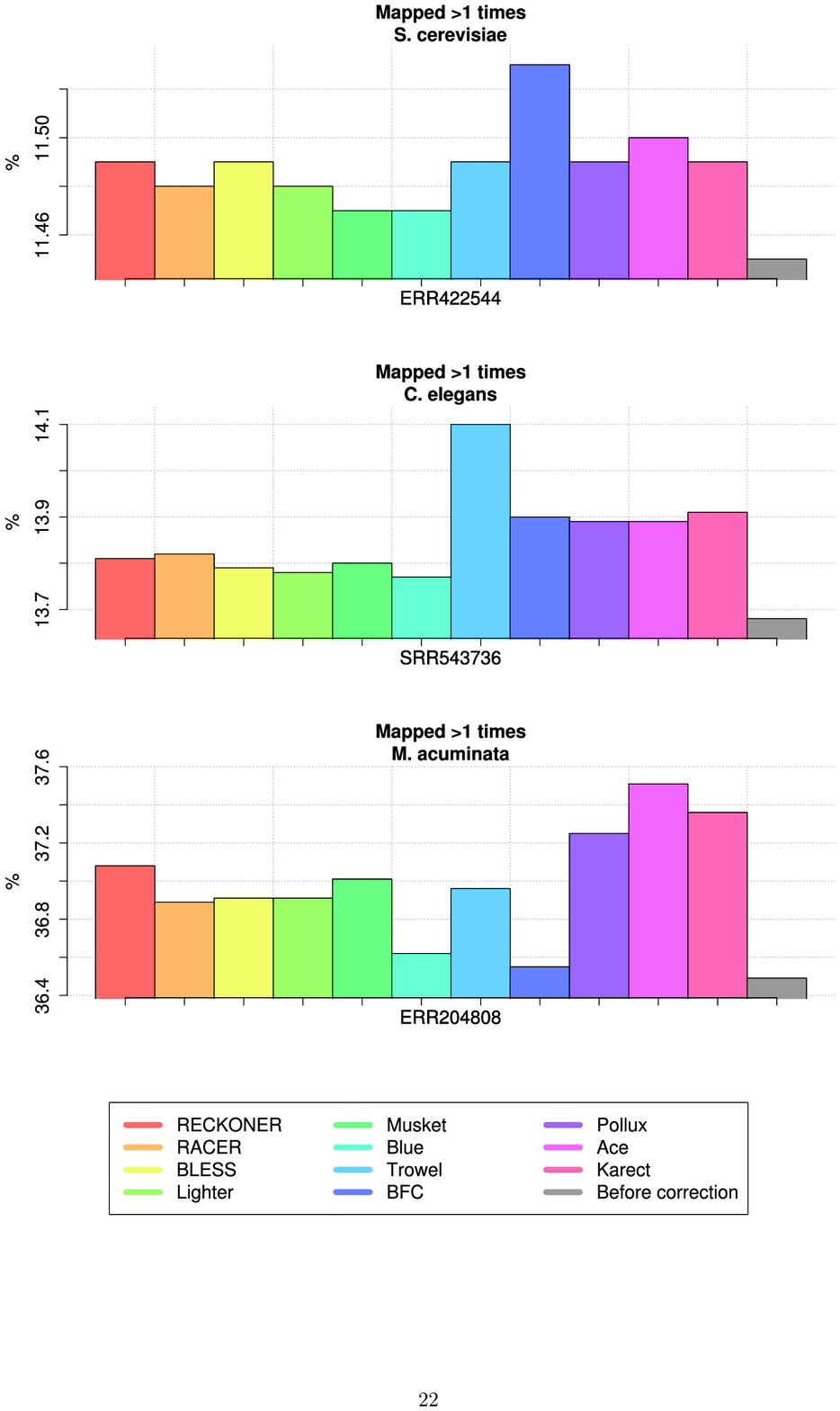}
\end{figure}
\clearpage
\begin{figure}[p]
\includegraphics{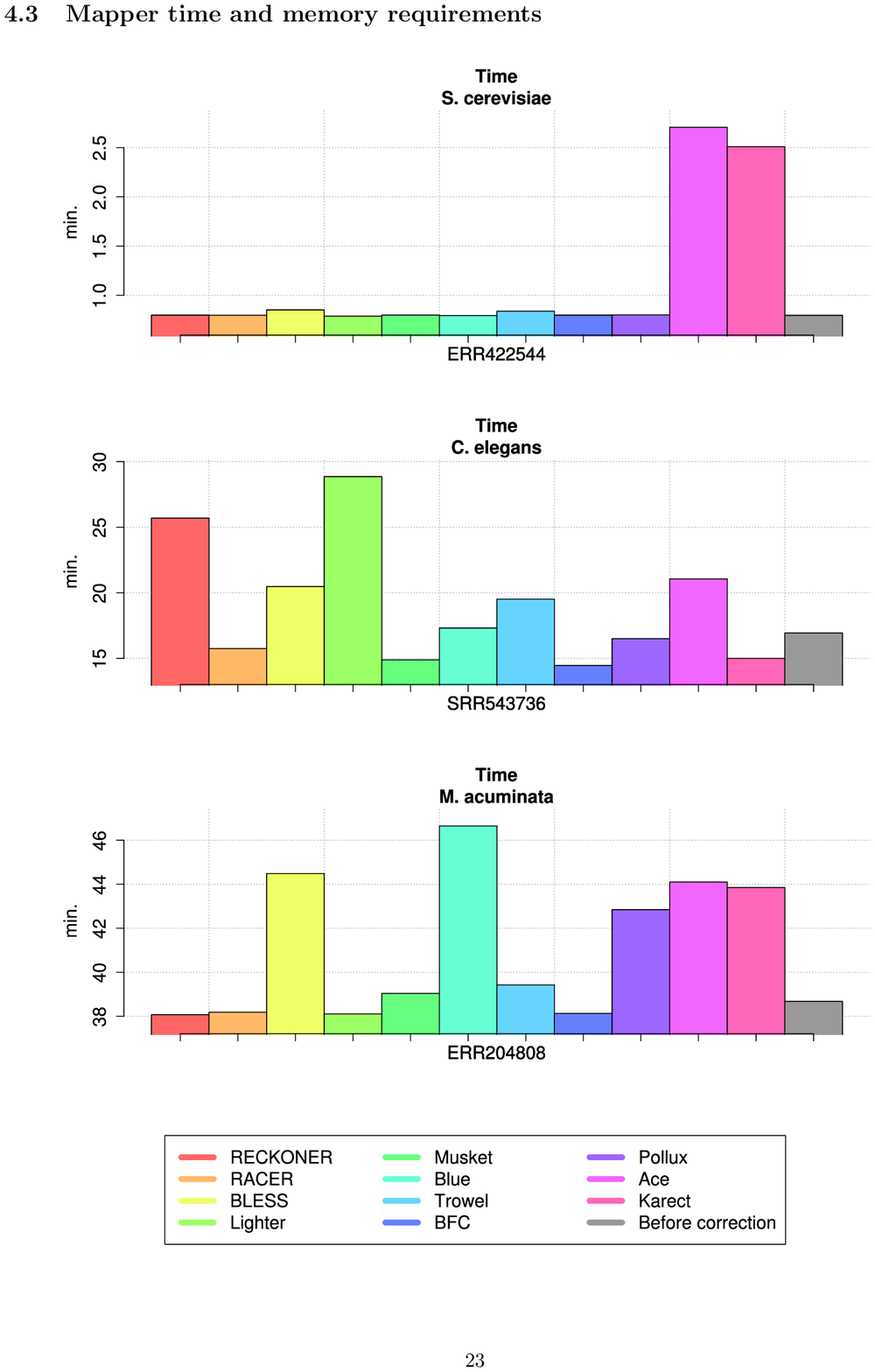}
\end{figure}
\clearpage
\begin{figure}[p]
\includegraphics{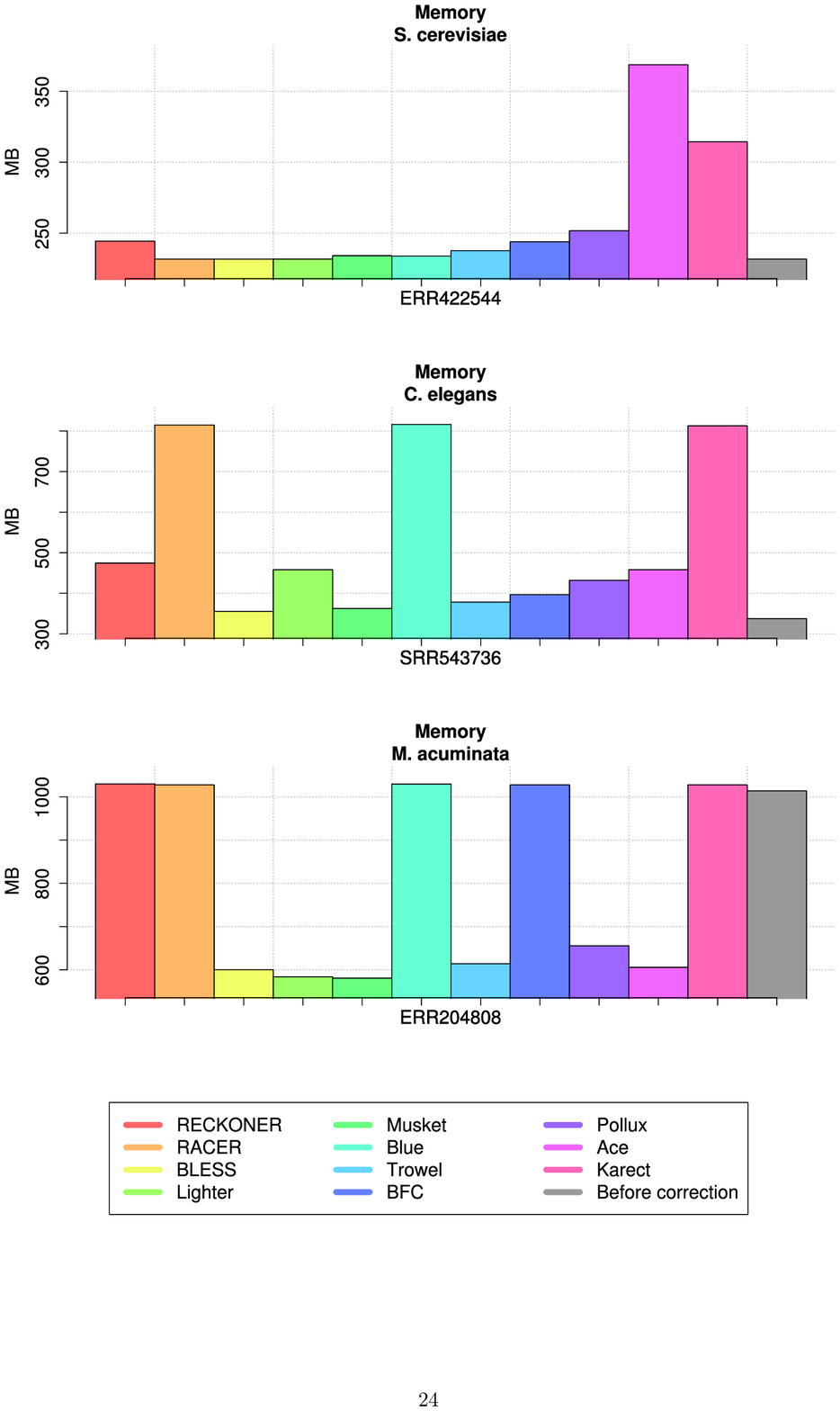}
\end{figure}
\end{document}